%% file: qfthep.tex
\documentclass[a4paper,10pt]{article}

\usepackage{contribution}

\input{econfmacros}
\input{contributionmacros}

\begin{document} 

\input{contribution}

\end{document}

%% file: econfmacros.tex
\newcommand{\weblink}[2][]{%
    \ifthenelse{\equal{#1}{}}%
    {\textnormal{\url{#2}}}%
    {\textnormal{\href{#2}{#1}}}%
}

\def\beq{\begin{equation}}
\def\eeq#1{\label{#1}\end{equation}}
\def\eeqn{\end{equation}}

\def\beqa{\begin{eqnarray}}
\def\eeqa#1{\label{#1}\end{eqnarray}}
\def\eeqan{\end{eqnarray}}

\let\bar=\overbar

\def\Dslash{\not{\hbox{\kern-4pt $D$}}}
\def\dslash{\not{\hbox{\kern-2pt $\del$}}}

\def\msb{{\bar{\ssstyle M \kern -1pt S}}}

%% file: contributionmacros.tex
\newcommand{\contribution}[7][]{%
  \clearpage
  \thispagestyle{plain}

  \ifthenelse{\equal{#1}{}}
  {\hypersetup{pdftitle={#2}}}
  {\hypersetup{pdftitle={#1}}}
  \hypersetup{pdfauthor={{#3} {#4}}}
  {\centering\normalfont\LARGE\bfseries\sffamily #2 \par\nobreak}
  \lhead{}
  \chead{%
    \textit{\footnotesize XXIInd International Workshop ``High-Energy Physics and Quantum Field Theory'', 
June 24 -- July 1, 2015, Samara, Russia}%
  }
  \rhead{}
  \bigskip
  \begin{center}
    {#3} {#4}\ifthenelse{\equal{#6}{}}{}{\footnote{\weblink[#6]{mailto:#6}}}
    \ifthenelse{\equal{#7}{}}{}{#7} \\
    \textit{#5}
  \end{center}
  \bigskip
}

\renewcommand{\abstract}[1]{%
  \begin{center}
    \begin{minipage}{0.85\textwidth}
      \begin{footnotesize}
        #1
      \end{footnotesize}
    \end{minipage}
  \end{center}
  \bigskip
}

%% file: contribution.tex
%
%
%
%
%
%
{  


%

\contribution[Recent ATLAS results and preparations for Run~2]  
{Recent ATLAS results}  
{Gabriella}{P\'asztor\footnote{
Now at MTA -- ELTE Lend\"ulet CMS Particle and Nuclear Physics Group, E\"otv\"os
Lor\'and University, Budapest, Hungary. E-mail: Gabriella.Pasztor@cern.ch}  
}  
{Carleton University, Ottawa, Canada}  
{}
{on behalf of the ATLAS Collaboration}  
%

\abstract{%

The CERN LHC delivered 25~fb$^{-1}$ of proton-proton collision data in 2011$-$2012 at $\sqrt{s} = 7 - 8$~TeV centre-of-mass energy to the ATLAS detector. These Run 1 data were used to discover the Higgs boson and measure its properties as well as to perform numerous other tests of the Standard Model via precision measurements and searches for New Physics. 
In preparation for the Run 2 data taking at $\sqrt{s} = 13$~TeV, the collaboration embarked on an upgrade program during the long LHC machine shutdown in 2013$-$2015. 
The early 2015 data set was then used to promptly recommission the apparatus allowing the first
physics results to appear already in the summer of 2015.
After briefly reviewing the upgrade and the improved performance of the detector, this paper concentrates on the final Run~1 results on Higgs, W and Z boson, electroweak multi-boson and top quark production as well as on searches for supersymmetry and other new phenomena.
}
%


\section{Introduction}

The Standard Model (SM) of particle physics stands up very well to scrutiny at high-energy particle collisions and describes with great precision the observed phenomena. However, both experimental hints (such as the presence of dark matter and dark energy in the Universe or the small but non-zero masses of neutrinos) as well as theoretical considerations (such as the desire to formulate a common theoretical framework that includes gravity and the three gauge interactions of the SM) call for New Physics which could manifest itself at the TeV energy scale in proton-proton collisions at the Large Hadron Collider (LHC) at CERN. The ATLAS Collaboration has a very rich physics program to test the SM, to measure its parameters and to study in detail the observed elementary particles, among them the Higgs boson discovered in 2012 at the LHC Run~1 data at centre-of-mass energies of $\sqrt{s} = 7 - 8$~TeV. The quest to explore New Physics beyond the SM relies both on these precision measurements and on direct searches for new phenomena. The increased LHC collision energy to $\sqrt{s} = 13$~TeV in 2015 opens up new territories for research in Run~2, bringing the hope of new discoveries. 

To fully exploit the new LHC data an upgrade program was launched to enhance the detector capabilities. After a brief summary of the improved experimental performance, I review a selection of the final Run~1 measurements that set the schene for the much awaited Run~2 results. 
 
\section{ATLAS detector upgrades for LHC Run~2}

The ATLAS detector~\cite{ATLAS-Detector} is a multi-purpose apparatus at the LHC. It consists of four main subsystems: an inner tracking system, electromagnetic (EM) and hadronic calorimeters, and a muon spectrometer. The inner detector provides tracking information from pixel and silicon microstrip detectors in the pseudorapidity range\footnote{
ATLAS uses a right-handed coordinate system with its origin at the nominal interaction point (IP) in the centre of the detector and the $z$-axis along the beam pipe. The $x$-axis points from the IP to the centre of the LHC ring, and the $y$-axis points upward. Cylindrical coordinates $(r,\phi)$ are used in the transverse plane, $\phi$ being the azimuthal angle around the $z$-axis. The pseudorapidity is defined in terms of the polar angle $\theta$ as $\eta=-\ln\tan(\theta/2)$.} 
of $|\eta|<2.5$ and from a transition radiation tracker (TRT) covering $|\eta|<2.0$, all immersed in a 2 T magnetic field provided by a thin superconducting solenoid. The finely segmented EM sampling calorimeter uses lead and liquid argon (LAr) and is divided into barrel ($|\eta|<1.475$) and endcap ($1.375<|\eta|<3.2$) regions. Hadron calorimetry is provided by a steel / scintillator-tile calorimeter, segmented into three barrel structures in the range of $|\eta|<1.7$, and two copper / LAr hadronic endcap calorimeters that cover the region of $1.5<|\eta|<3.2$.
The solid angle coverage is completed with forward copper / LAr and tungsten / LAr calorimeter modules, optimised for EM and hadronic measurements respectively, and covering the region of $3.1<|\eta|<4.9$. The muon spectrometer measures the deflection of muon tracks in the range of $|\eta|<2.7$ using multiple layers of high-precision tracking detectors (Monitored Drift Tubes and Cathode Strip Chambers) located in toroidal magnetic fields of approximately 0.5 T and 1 T in the central and endcap regions of ATLAS, respectively. The muon spectrometer is also instrumented with separate trigger layers of Resistive Plate Chambers and Thin Gas Chambers covering $|\eta|<2.4$.

A trigger system~\cite{ATLAS-Trigger} reduces the event rate to be recorded to about 1 kHz from the LHC beam crossing rate of 40 MHz. It is based on the Region-of-Interest concept in which the software-based high-level trigger (HLT) reconstruction is seeded by the level-1 (L1) objects provided by the hardware trigger with less than 2.5 $\mu$s latency. In Run~1, the HLT was divided to a level-2 (L2) and an event filter (EF) step.

The long shutdown of the LHC allowed to consolidate and to partially upgrade the detector and its trigger and data acquisition system. 

Most notably a new pixel detector barrel layer (Insertable B-layer, IBL)~\cite{ATLAS-IBL} was installed close to the beam pipe at a radius of 33 mm.
It employs two different sensor technologies: planar and 3D, both with a pixel size of $50 \times 250 \mu$m. The modules are supported with lightweight carbon foam structures and are cooled by a CO$_2$-based system. The new layer provides robustness against irreparable failures and high occupancy at the higher expected instantaneous luminosities of Run~2 as well as improves tracking precision. 
For example, the transverse and longitudinal impact parameter resolutions improve with the addition of the IBL by up to 40\%, especially at low transverse momentum.



To consolidate the detector, the muon coverage was completed by installing new chambers in the Extended Endcap, as well as new Diamond Beam Monitors and LUCIDs (Luminosity measuring Cherenkov Integrating Detectors) were added. The Beam Conditions Monitors were upgraded and repairs were performed on several systems (TRT, LAr and Tile calorimeters...). The infrastructure was also improved by installing a new beam pipe, muon chamber shielding and pixel detector services. 

The increased energy and luminosity of the LHC in Run~2 necessitated the upgrade of the trigger system~\cite{ATLAS-TriggerUpgrade} to keep event rates under control while maintaining high efficiencies for interesting processes. The first part of the ambitious upgrade program was successfully completed during 2013 $-$ 2015. The maximal L1 output rate was increased from 75 kHz to 100 kHz. A new Central Trigger Processor was installed that accepts inputs from a new L1 Topological Trigger Processor. The installation and commissioning of the new hardware Fast Tracker (FTK)~\cite{ATLAS-FTK} has also started during LS1. The FTK operates at the L1 output rate of 100 kHz providing tracks in 100 ms to the HLT. The full FTK system is expected to be operational by the end of 2016. 

The L2 and EF farms were merged enabling common data preparation for fast and precision calculations in the HLT. The final HLT decision is made now on average in 0.2 s. The HLT rate was increased and is now limited to 1.1 - 1.5 kHz by storage capacity. 

The online and offline reconstruction software was also improved and a new data format and analysis framework introduced to optimise the usage of available resources. 

\section{ATLAS data samples}

During Run~1, the ATLAS detector collected high-quality proton-proton collision data corresponding to an integrated luminosity of $\int\cal{L}$ = 4.57~fb$^{-1}$ at $\sqrt{s} = 7$~TeV in 2011 with an average number of collisions per beam crossing of $< \mu > = 9.1$ and $\int\cal{L}$ = 20.3~fb$^{-1}$ at $\sqrt{s} = 8$~TeV in 2012 with $< \mu > = 20.7$. The efficiency to record the delivered LHC data was more than 93\%. Almost 95\% of the recorded events also passed stringent quality criteria and were thus used for physics analyses.

\begin{figure}[!ht]
\centering
\includegraphics[width=8cm]{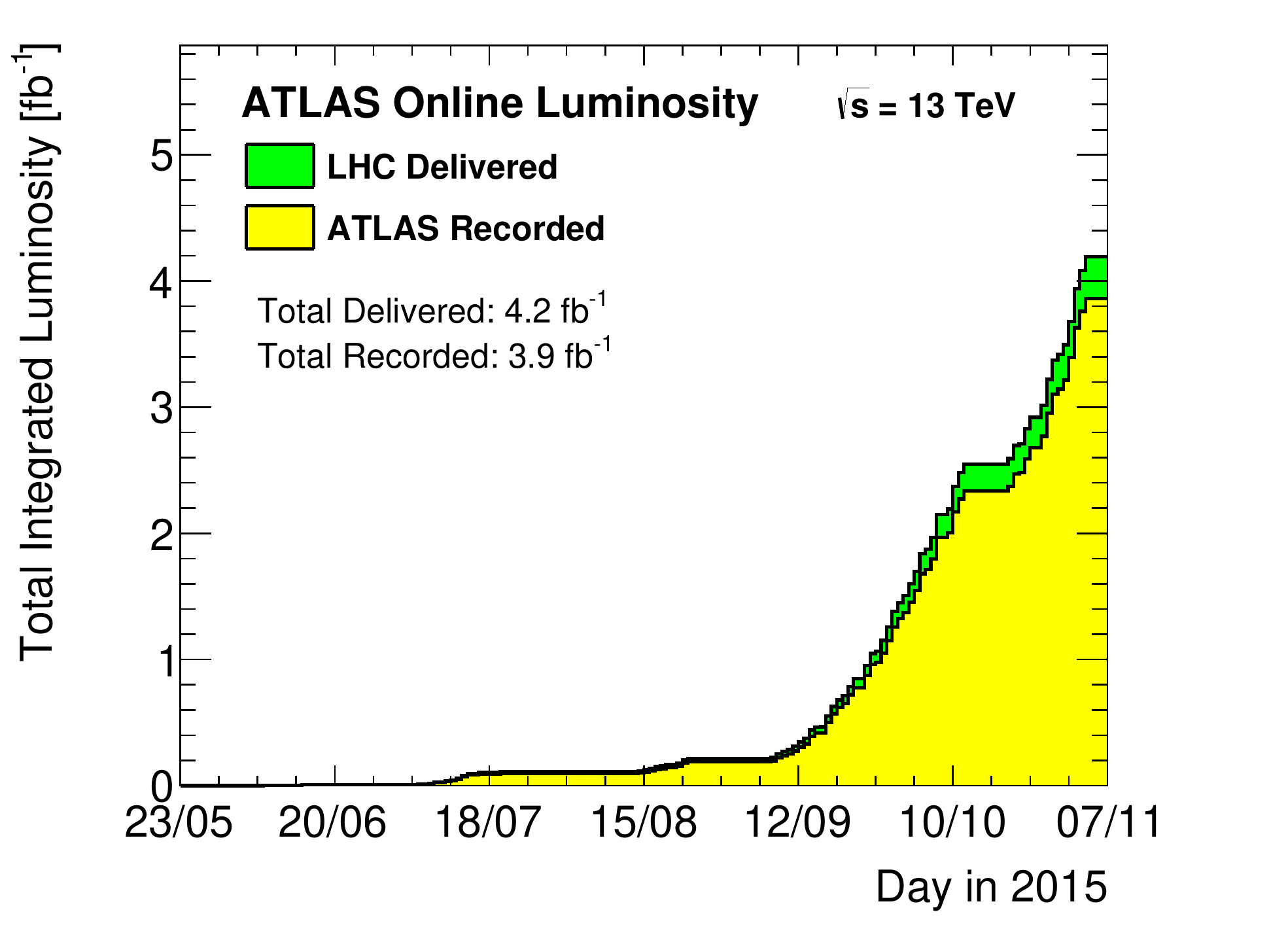} 
\includegraphics[width=8cm]{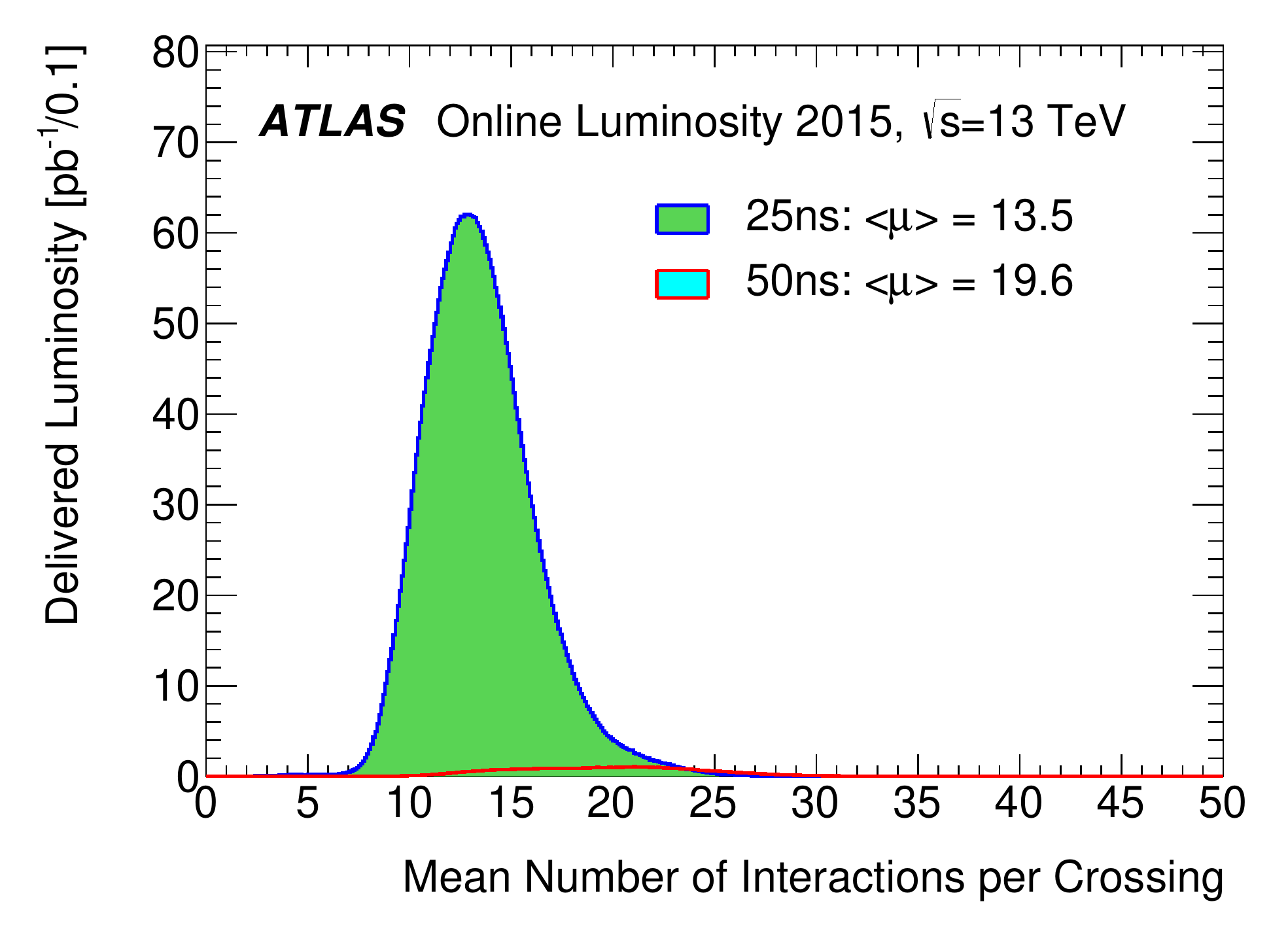} 
\caption{(left) Cumulative luminosity versus time delivered to (green) and recorded by ATLAS (yellow) during stable beams for pp collisions at 13~TeV centre-of-mass energy in 2015. 
(right) The luminosity-weighted distribution of the mean number of interactions per crossing for the 2015 pp collision data recorded from 3 June - 3 November at 13~TeV centre-of-mass energy. 
Taken from Ref.~\cite{ATLAS-Lumi}.
\label{fig:Lumi2015}}
\end{figure}

In 2015, data taking efficiency stayed high at 92\% as shown in Figure~\ref{fig:Lumi2015}. The data quality efficiency reached 93\% (87\% if requiring the seamless operation of the new IBL). Most of the data were taken with 25 ns bunch spacing (the remaining with 50 ns) and a mean number of interactions per beam crossing of $< \mu > = 13.5$. From the total of $\int\cal{L}$ = 3.9~fb$^{-1}$ of 2015 data~\cite{ATLAS-Lumi}, at the time of the QFTHEP conference only $\int\cal{L}$ = 7.9~pb$^{-1}$ were available which were used for detector commissioning.

The 2015 data were collected by about 1500 HLT selections seeded by about 400 L1 items. The primary single electron trigger threshold was kept at 24~GeV, while the single muon threshold at 20~GeV. The lowest unprescaled missing transverse energy trigger operated with a 70~GeV threshold.



\section{Standard Model physics}

So far precision measurements at the LHC show excellent agreement with theoretical predictions through 14 orders of magnitude in cross-section, from the total pp interaction rate to the rate of the rarest processes measured such as electroweak (EW) vector boson scattering (VBS), as shown\footnote{
Results shown in this paper are those that were available at the time of the QFTHEP 2015 conference. For updates see Ref.~\cite{ATLAS-Results}.} 
in Figure~\ref{fig:SMSummary} from Ref.~\cite{ATLAS-SMSummary}. The discovery of the Higgs boson in Run~1 not only completed the SM but also opened up a new window to further test its boundaries. 


\begin{figure}[!ht]
\centering
\includegraphics[width=16cm]{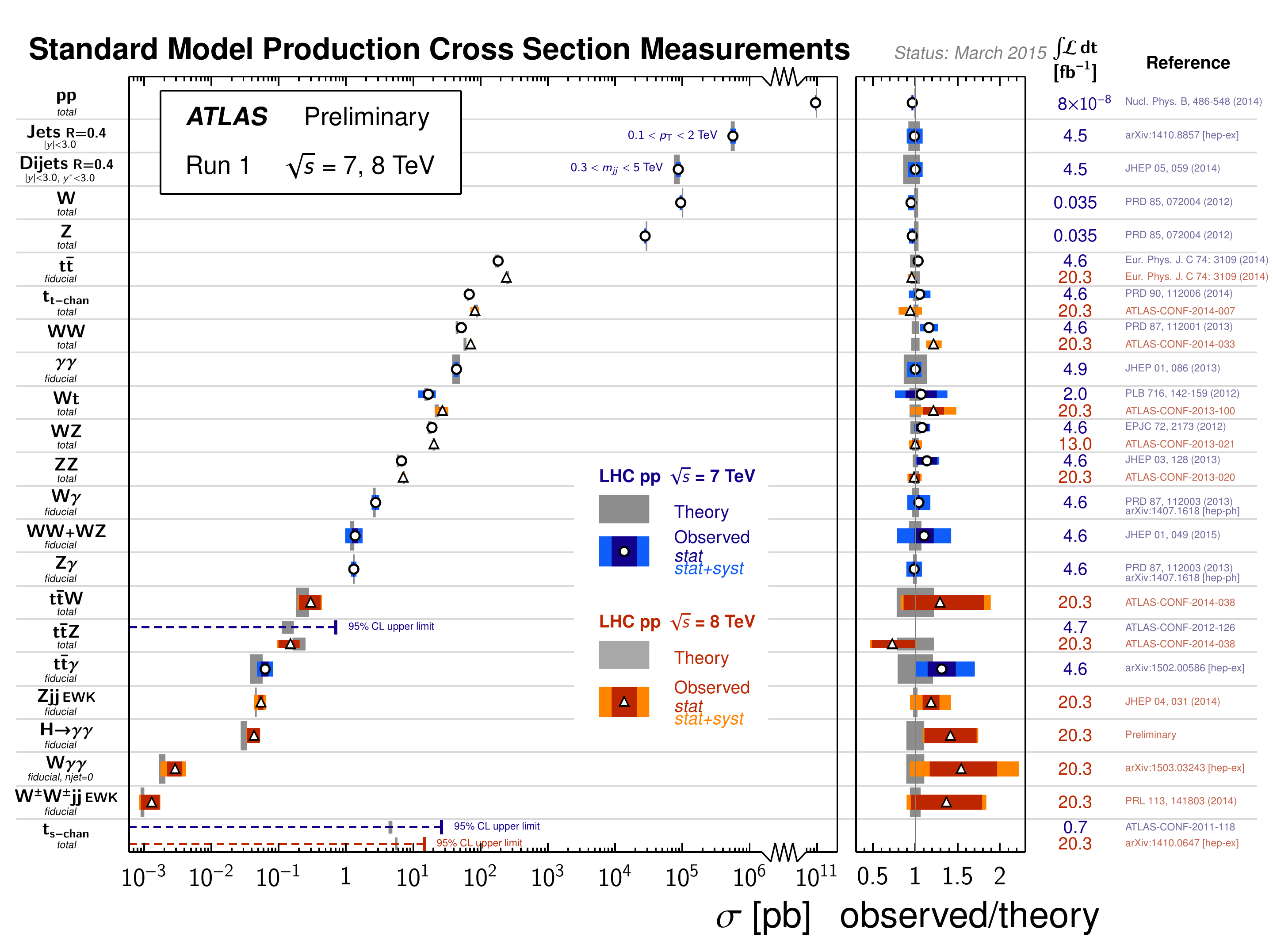}
\caption{
Summary of Standard Model total and fiducial cross-section measurements, corrected for leptonic branching fractions, compared to the corresponding theoretical expectations. All theoretical expectations were calculated at next-to-leading order or higher. 
Taken from Ref.~\cite{ATLAS-SMSummary}.
\label{fig:SMSummary}}
\end{figure}

\subsection{Higgs boson physics}

Since its discovery in 2012 by the ATLAS and CMS collaborations, the Higgs boson, the most unique particle of the SM was studied in great detail. 

The SM Higgs boson is produced dominantly via gluon fusion (ggF, 86\% of the total rate at $\sqrt{s}=7-13$~TeV) at the LHC. In the coupling measurements, however, the vector boson fusion (VBF, 7\%), the associated production with a weak vector boson (VH, 5\%) and the associated production with a top pair (ttH, $\leq 1$\%) play an essential role. These production modes can be recognised by requiring the reconstructed Higgs boson to be accompanied by two forward jets with little other hadronic activity, a W/Z boson or two top quarks, respectively. 

The Higgs boson coupling to fermions is proportional to the fermion mass, while its coupling to weak bosons is proportional to the mass squared. The best "discovery" channel is $H \rightarrow \gamma\gamma$ (branching ratio BR = 0.23\%) as it has good mass resolution and reasonable rates. The "golden channel" of $H \rightarrow ZZ^* \rightarrow \ell^+\ell^-\ell^+\ell^-$ (BR = 0.012\%) has the best signal-to-background ratio, excellent mass resolution and a fully reconstructable final state that provides unique angular information for spin-parity measurements. The last bosonic decay mode $H \rightarrow WW^* \rightarrow \ell^+\nu\ell^-\nu$ (BR = 1.0\%) provides high statistics but due to the final state neutrinos it has low resolution and a background that is challenging to estimate precisely. Fermionic channels are even more challenging. The best one is $H \rightarrow \tau^+ \tau^-$ (BR = 6.3\%). The dominant contribution to the total width comes from  $H \rightarrow b\bar b$ (BR = 57.5\%). This is also the best channel to measure VH production. Due to the high background this decay is not accessible in ggF. 
Even in the final Run 1 data, the observed significance of $H \rightarrow b\bar b$ process does not reach the 3$\sigma$ level.
Rare decays $H \rightarrow Z\gamma \rightarrow \ell^+\ell^-\gamma$ (BR = 0.01\%) and $H \rightarrow \mu^+\mu^-$ (BR = 0.02\%) are also searched for.
  
Higgs boson measurements require excellent object (electron, muon, tau, jet, b-jet and missing transverse energy) reconstruction performance and sophisticated analysis techniques to discriminate against the overwhelming backgrounds.

The SM does not predict the mass of the Higgs boson. However for a given mass the model gives precise predictions for the production cross-sections and decay rates. It is thus important to measure the mass with high precision. This is done using the high-resolution channels  $H \rightarrow \gamma\gamma$ and $H \rightarrow ZZ^* \rightarrow \ell^+\ell^-\ell^+\ell^-$ using a global fit to the data. With the final electromagnetic energy and muon transverse momentum scale calibrations, the combined ATLAS and CMS result is $m_H = 125.09 \pm 0.21 \mathrm{(stat.)} \pm 0.11 \mathrm{(0.11)}$~GeV~\cite{LHC-HiggsMass}, its uncertainty dominated by the available data statistics. 

The coupling measurements~\cite{ATLAS-HiggsCouplings} rely on the experimental separation of the different production and decay modes using their specific characteristics. The global fit to the event counts in the various phase-space regions take into account the background contributions. There are more than 1000 nuisance parameters corresponding to the different sources of systematic uncertainties in these fits. All results assume a single SM-like (CP-even scalar) Higgs boson with the tensor structure of the SM interactions and a small width. 

Figure~\ref{fig:HiggsCoupling} shows some of the results.
As the measured rates are only sensitive to the cross-section times the branching ratios, the most model-independent fit constrains ratios of cross-sections and branching ratios. The absolute normalisation is done to the $\sigma(gg \rightarrow H \rightarrow ZZ^*)$ cross-section chosen due to its smallest expected systematic uncertainty. The most discrepant values are related to the higher than expected ttH cross-section and the lower than expected rate for $H \rightarrow b\bar b$. These are also among the least precisely measured quantities.

\begin{figure}[!ht]
\centering
\hspace*{-0.15cm} \includegraphics[width=5.6cm, height=6cm, bb=0 -15 580 610]{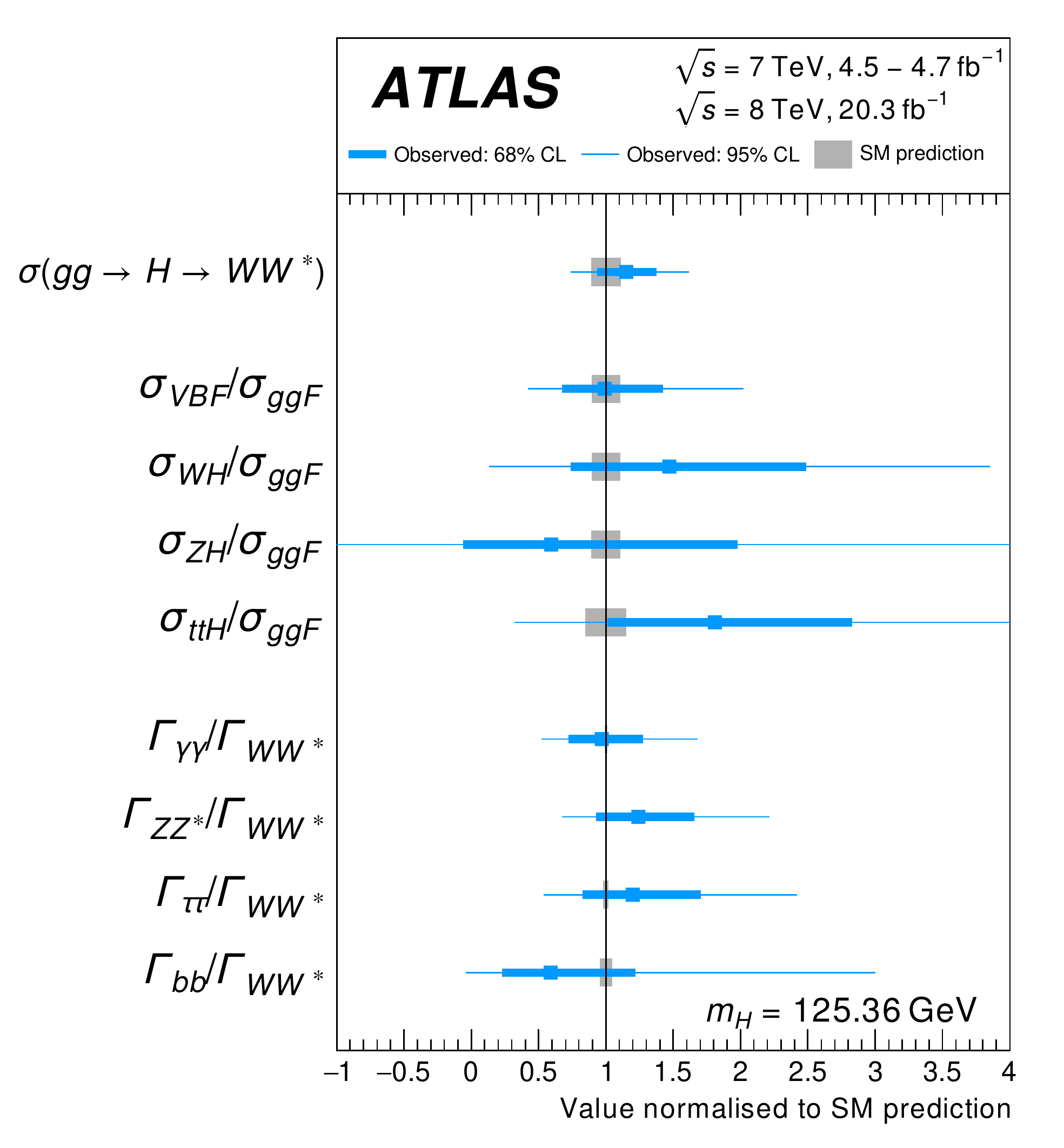} \hspace*{-0.4cm} \includegraphics[width=6cm, height=6.1cm]{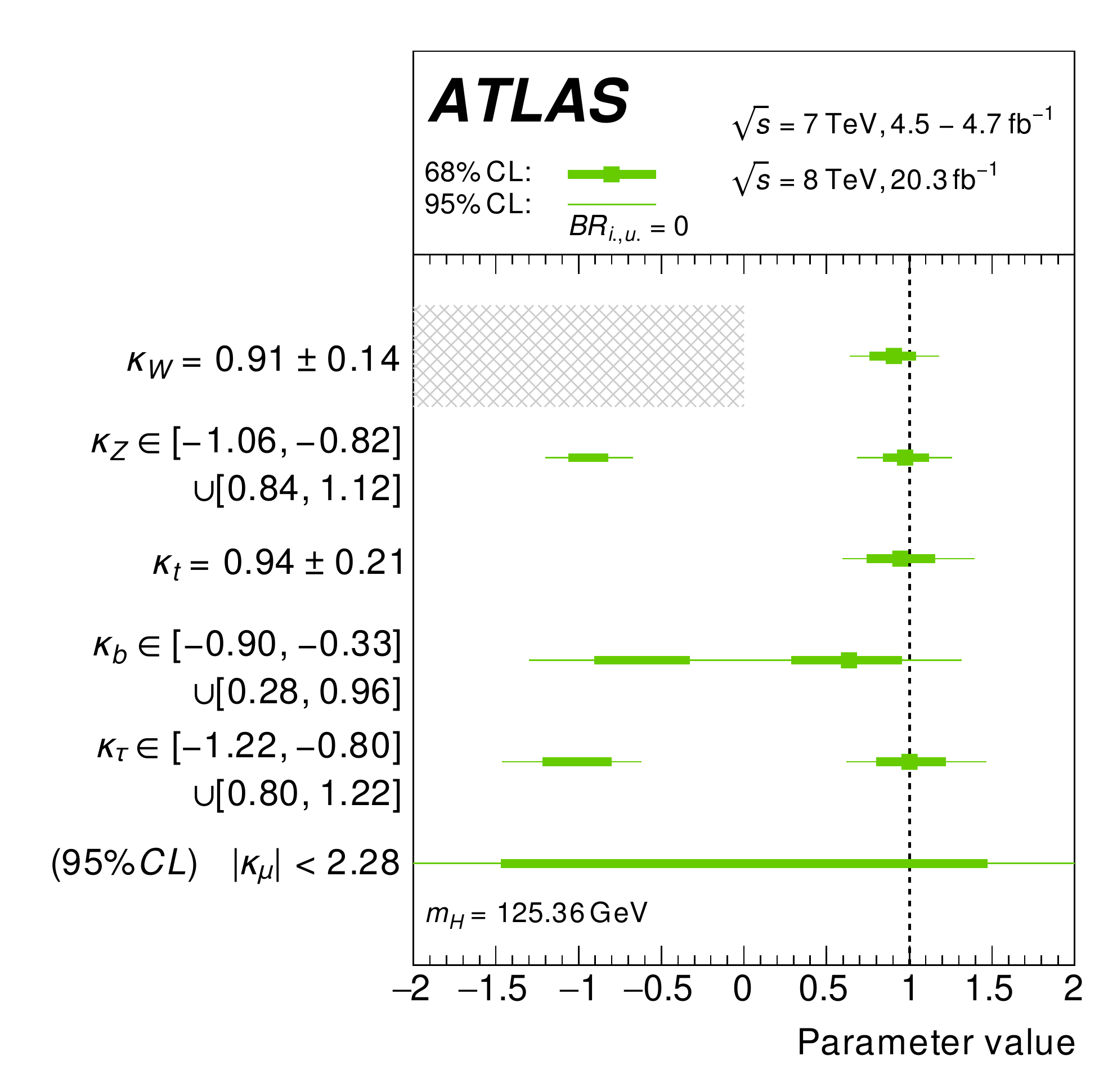} \hspace*{0.05cm} 
\includegraphics[width=5cm, height=4.6cm]{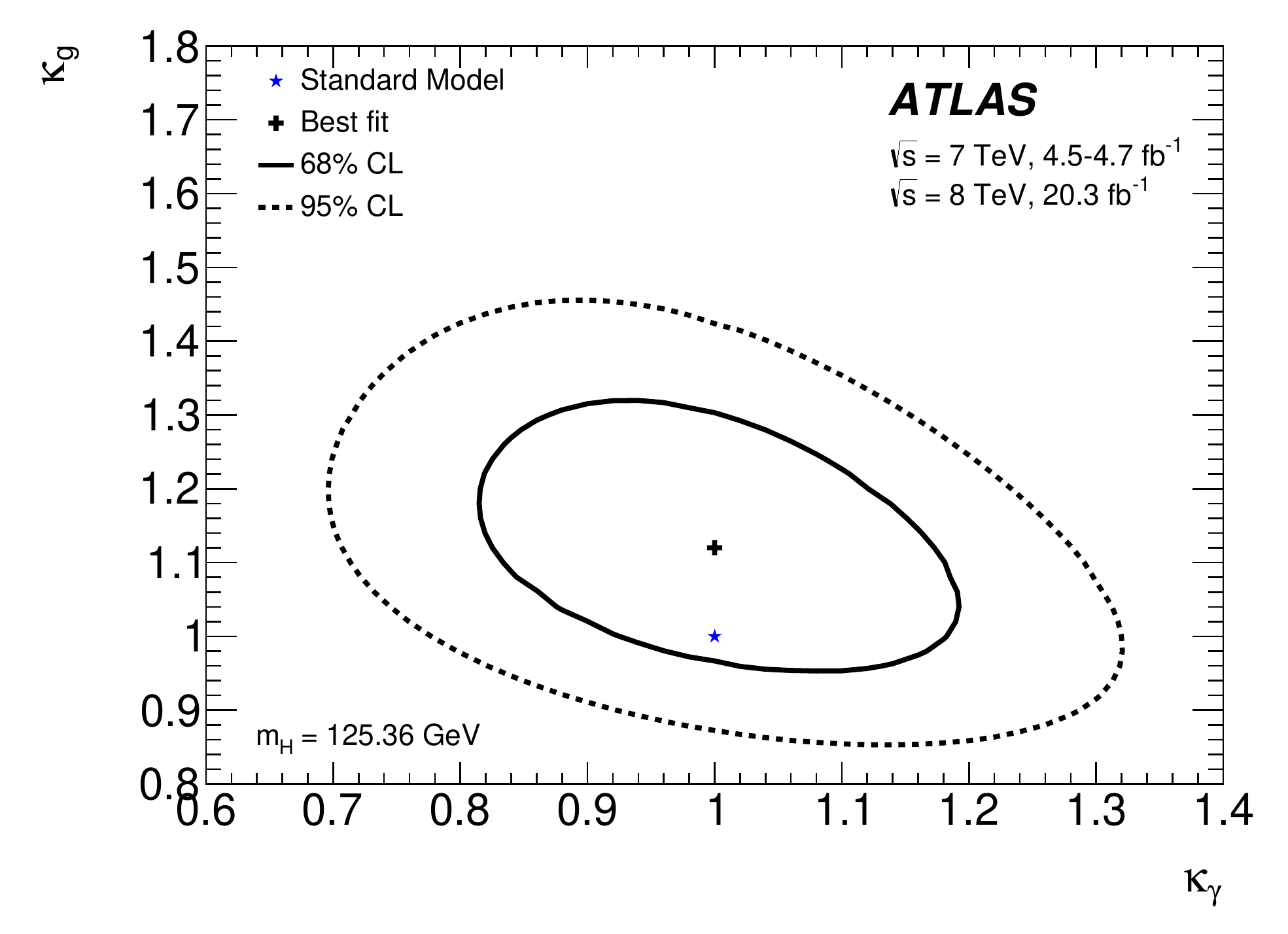} \hspace*{-0.15cm}
\caption{(left) Best-fit values of the $\sigma(H \rightarrow ZZ^*)$ cross-section and of ratios of cross-sections and branching ratios. (middle) Measurements of the coupling modifiers assuming no BSM contributions to the total width. (right) Constraints on BSM contributions to the loop coupling modifiers in a generic model where New Physics contributes only in the $gg \rightarrow H$ and $H \rightarrow \gamma\gamma$ loops. 
Taken from Ref.~\cite{ATLAS-HiggsCouplings}.
\label{fig:HiggsCoupling}}
\end{figure}

Various measurements are also performed with different model assumptions to determine coupling modifiers $\kappa_i = g_i / g_i^\mathrm{SM}$. A good agreement is observed with the SM predictions in all fits. An example is shown in the middle plot of Figure~\ref{fig:HiggsCoupling} where no beyond the Standard Model (BSM) contribution is allowed to the Higgs decay width. In a general fit that allows direct decays to beyond the SM particles, the branching ratio of these is constrained to be BR$_\mathrm{BSM} < 0.13$ at the 95\% CL, assuming $\kappa_V \leq 1$. On the right of Figure~\ref{fig:HiggsCoupling} the measurement if the loop coupling modifiers is shown when BSM contribution is only allowed in the $gg \rightarrow H$ and $H \rightarrow \gamma\gamma$ loops.

\begin{figure}[!ht]
\centering
\includegraphics[width=5.4cm]{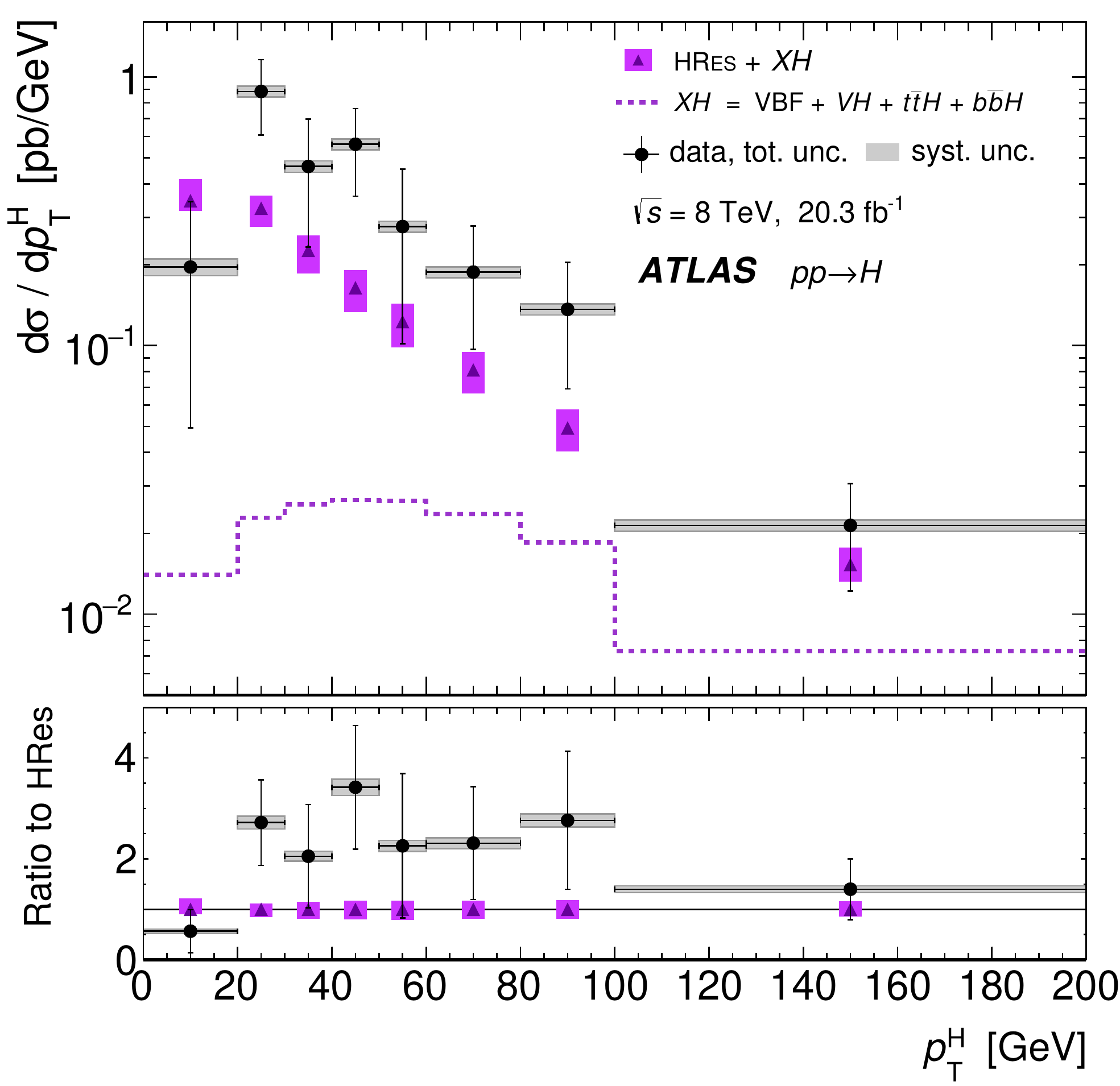}
\includegraphics[width=5.4cm]{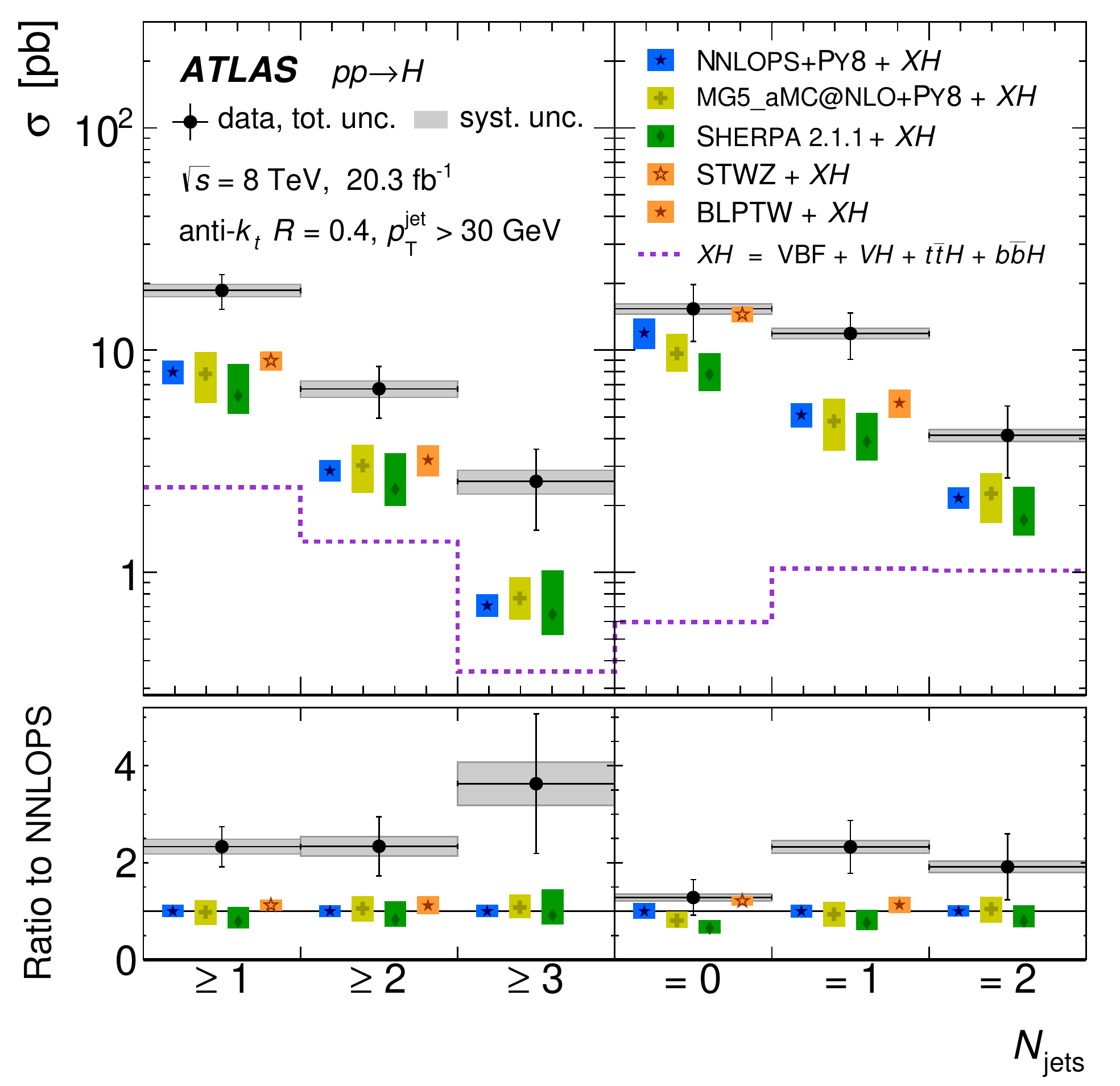}
\includegraphics[width=5.4cm]{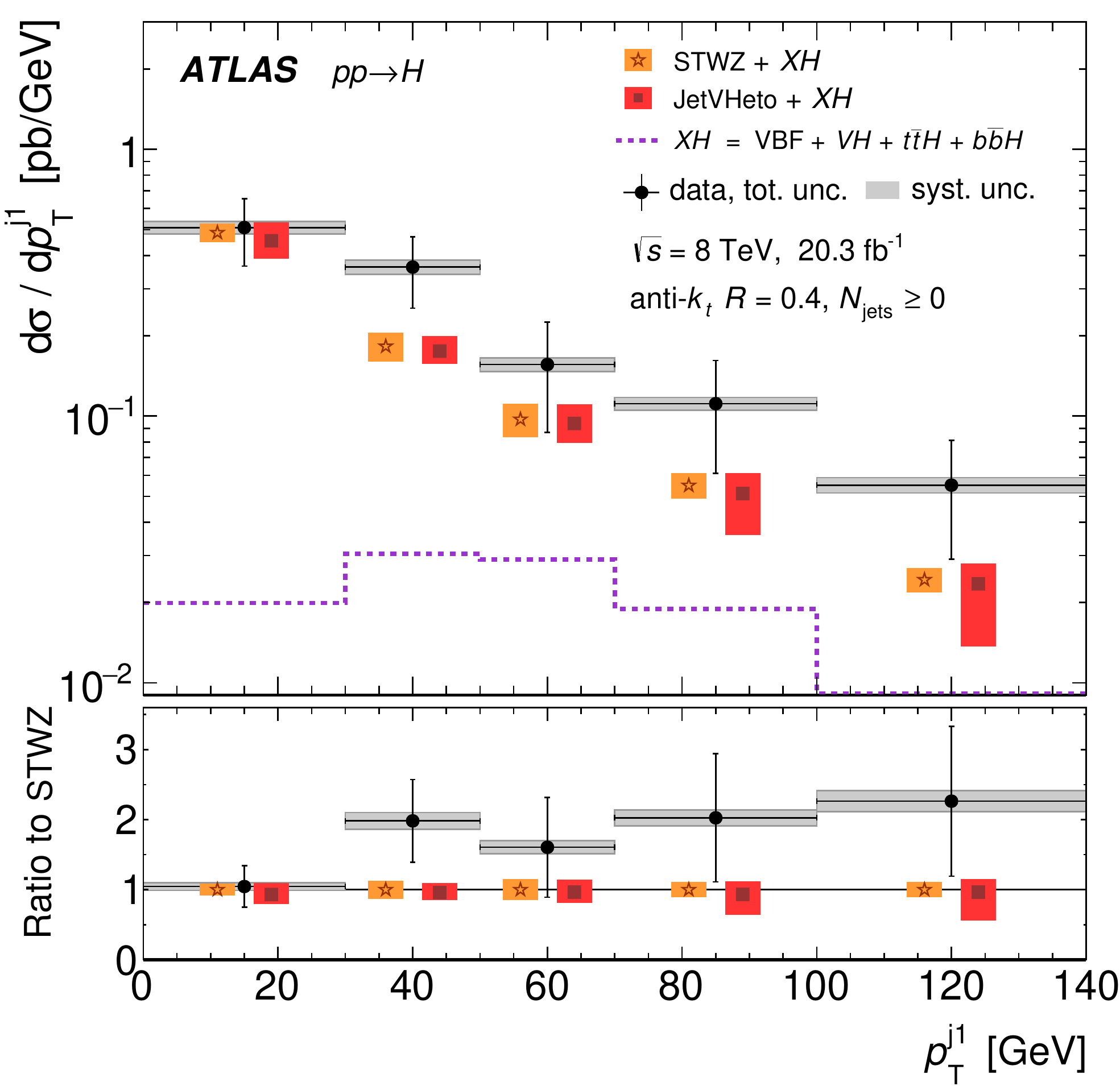} 
\caption{Observed fiducial and differential cross-sections from the combination of the $H \rightarrow \gamma\gamma$ and $H \rightarrow ZZ* \rightarrow 4\ell$ measurements compared to the SM prediction as a function of (left) the Higgs transverse momentum, (middle) the number of jets,  and (right) the transverse momentum of the most energetic jet.
Taken from Ref.~\cite{ATLAS-HiggsXsec}.
\label{fig:HiggsXsec}}
\end{figure}

Fiducial and differential cross-sections measurements~\cite{ATLAS-HiggsXsec} provide a model-independent study of the production and decay kinematics and allow comparisons to theoretical calculations both in and beyond the SM. They can test the modeling of different Higgs boson production mechanisms and are sensitive to New Physics. The combined results from the $H \rightarrow \gamma\gamma$ and $H \rightarrow ZZ* \rightarrow 4\ell$ measurements are shown in Figure~\ref{fig:HiggsXsec}. The Higgs boson transverse momentum distribution describes the kinematics and sensitive to the
perturbative QCD modeling of the dominant ggF production.
The jet distributions probe the theoretical description of partonic radiation in ggF production as well as the overall rate and modelling of jets in VBF and VH production. Jets produced in VBF, VH and ttH production tend to have higher transverse momenta than those coming from the ggF process.
All results are consistent with the SM.


Due to the limited statistics, spin-parity studies~\cite{ATLAS-HiggsSpin} use the measurements of the production and decay kinematics in the bosonic channels to test the compatibility of the data with the SM hypotheses against several alternative spin-parity scenarios. The best fit is always provided by the SM $J^P = 0^+$ hypothesis and all other tested models are excluded at more than 99.9\% CL. Recent measurements~\cite{ATLAS-HiggsSpin} also test the tensor structure of the interaction between the spin-0 boson and the SM vector bosons in an Effective Field Theory with a Lagrangian containing SM, BSM CP-even and BSM CP-odd contributions. Assuming only one BSM contribution at a time the measured BSM couplings are constrained and found to be compatible with zero as shown in Figure~\ref{fig:HiggsSpin}. 

\begin{figure}[!ht]
\centering
\includegraphics[width=7.8cm]{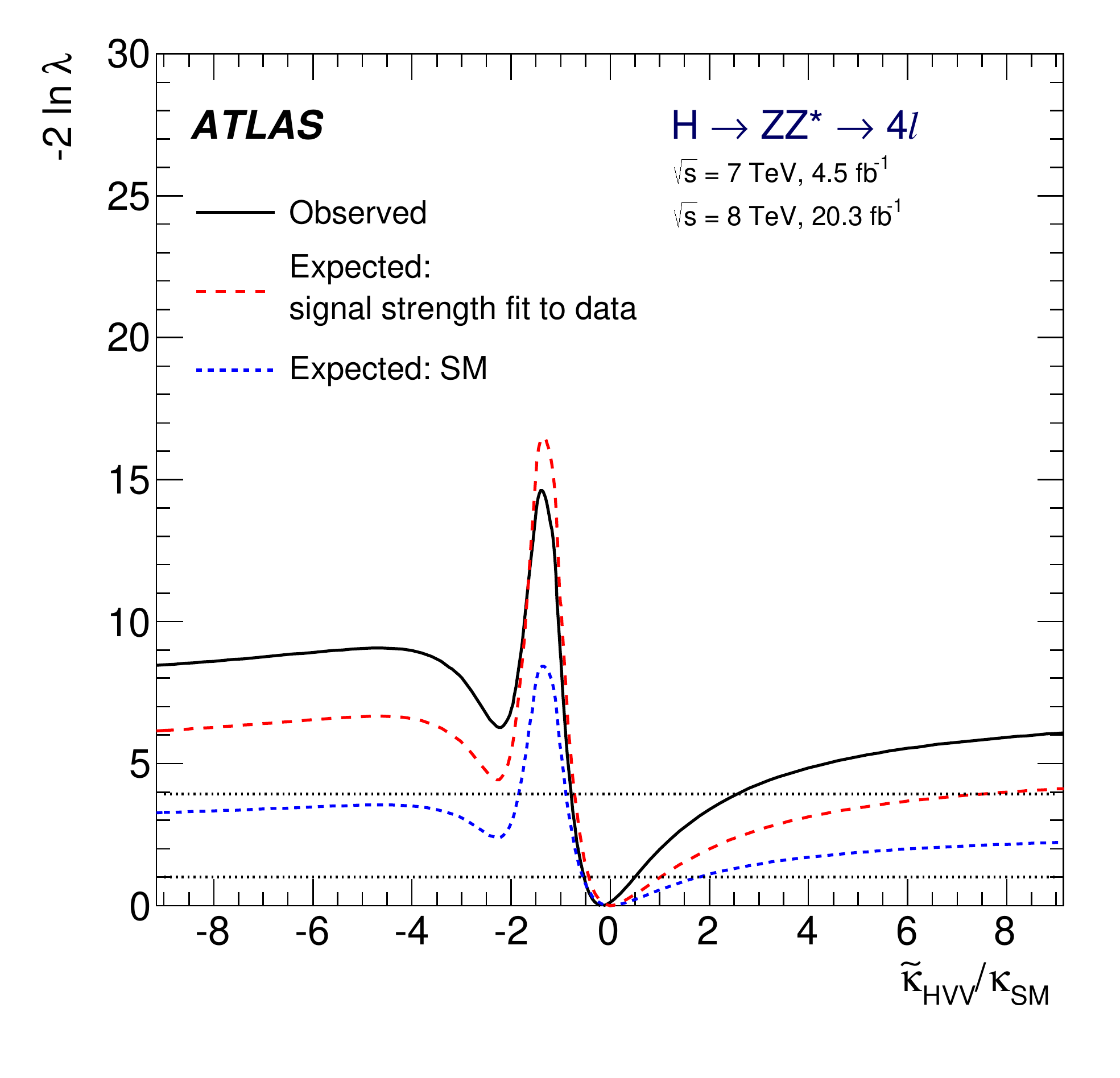}
\includegraphics[width=7.8cm]{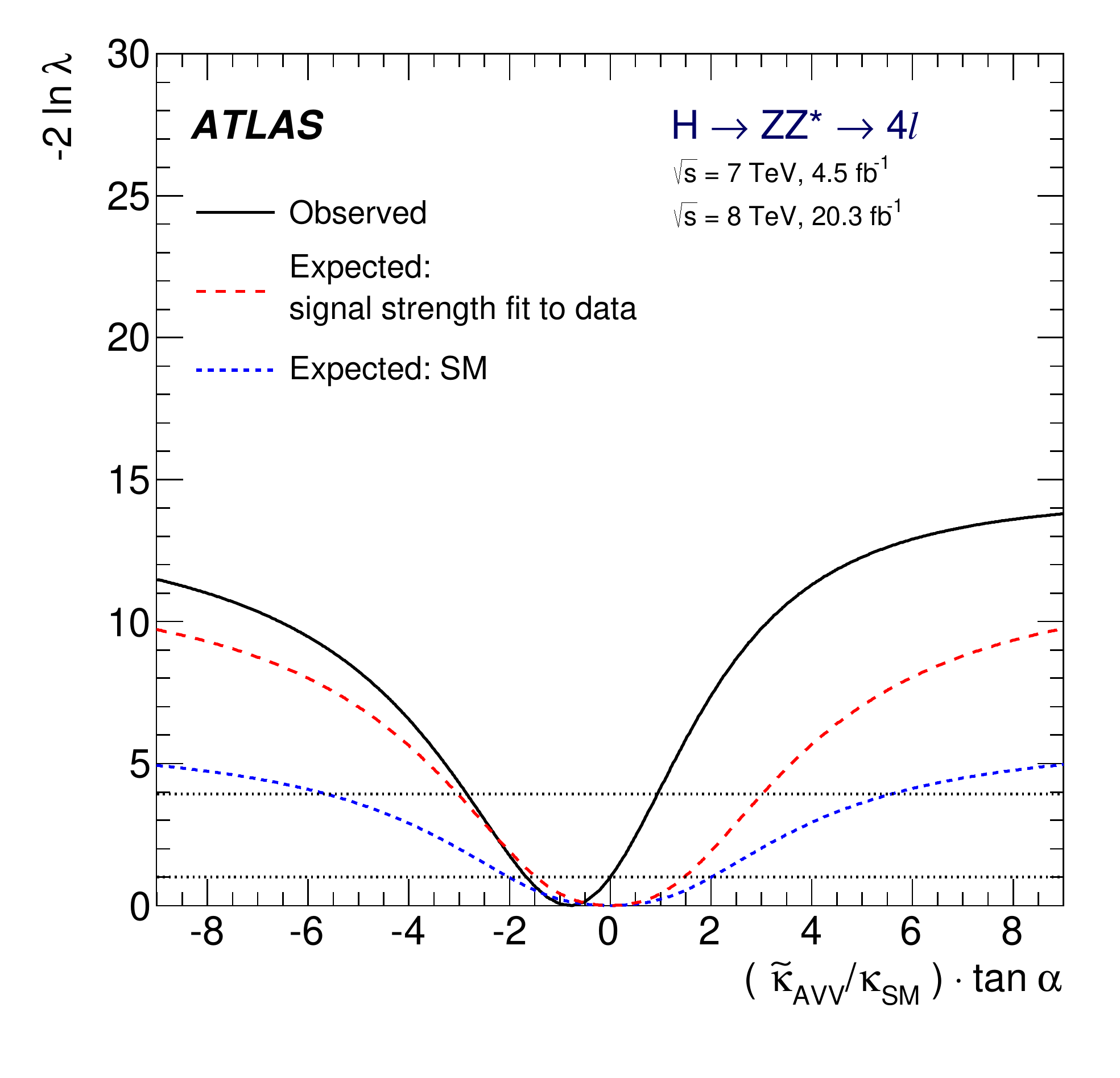} 
\caption{Constraints on BSM Higgs couplings for (left) CP-even and (right) CP-odd contributions, assuming the other contribution to be zero.
Taken from Ref.~\cite{ATLAS-HiggsSpin}.
\label{fig:HiggsSpin}}
\end{figure}

As the Higgs boson properties (spin, parity, couplings...) are measured with better and better precision and still show a high level of compatibility with theoretical prediction, the Standard Model of particle physics as an effective theory is completed with respect to its particle content. Nonetheless, the scalar sector could still provide the means to get information on the New Physics underlying the SM. Many searches thus aim to access exotic production and decay modes of the Higgs boson in a large number of alternative models as well as new Higgs states, so far with no positive result in Run 1 data.
However, before proceeding to discuss results on New Physics, one needs to step back to the traditional Standard Model measurements of W boson, Z boson and top quark production that laid down the bases on which all searches are built.

\section{W and Z boson production}

The study of the forward - backward asymmetry in the polar angle distribution of final state leptons in the $Z/\gamma^* \rightarrow \ell^+ \ell^-$ process~\cite{ATLAS-Afb} provides a measurement of the effective weak mixing angle (see Figure~\ref{fig:Afb}) as well as the as the asymmetry parameter $A_f$ for fermion $f$ directly related to its electroweak vector and axial-vector couplings. The measurement benefits from the ATLAS capability of identifying electrons in the forward region at $2.5 < |\eta| < 4.9$, with the central - forward di-electron channel providing the most precise measurement of $\sin^2\theta_\mathrm{eff}^\mathrm{lept}$. While the uncertainty of the combined Run~1 $\sqrt{s} = 7$~TeV measurement is an order of magnitude away from that of the world average, it is only about four times of that of the leading and discrepant LEP and SLC measurements. The dominant uncertainty comes from the parton distribution functions (PDFs) of the proton.

\begin{figure}[!h]
\centering
\includegraphics[width=5.8cm]{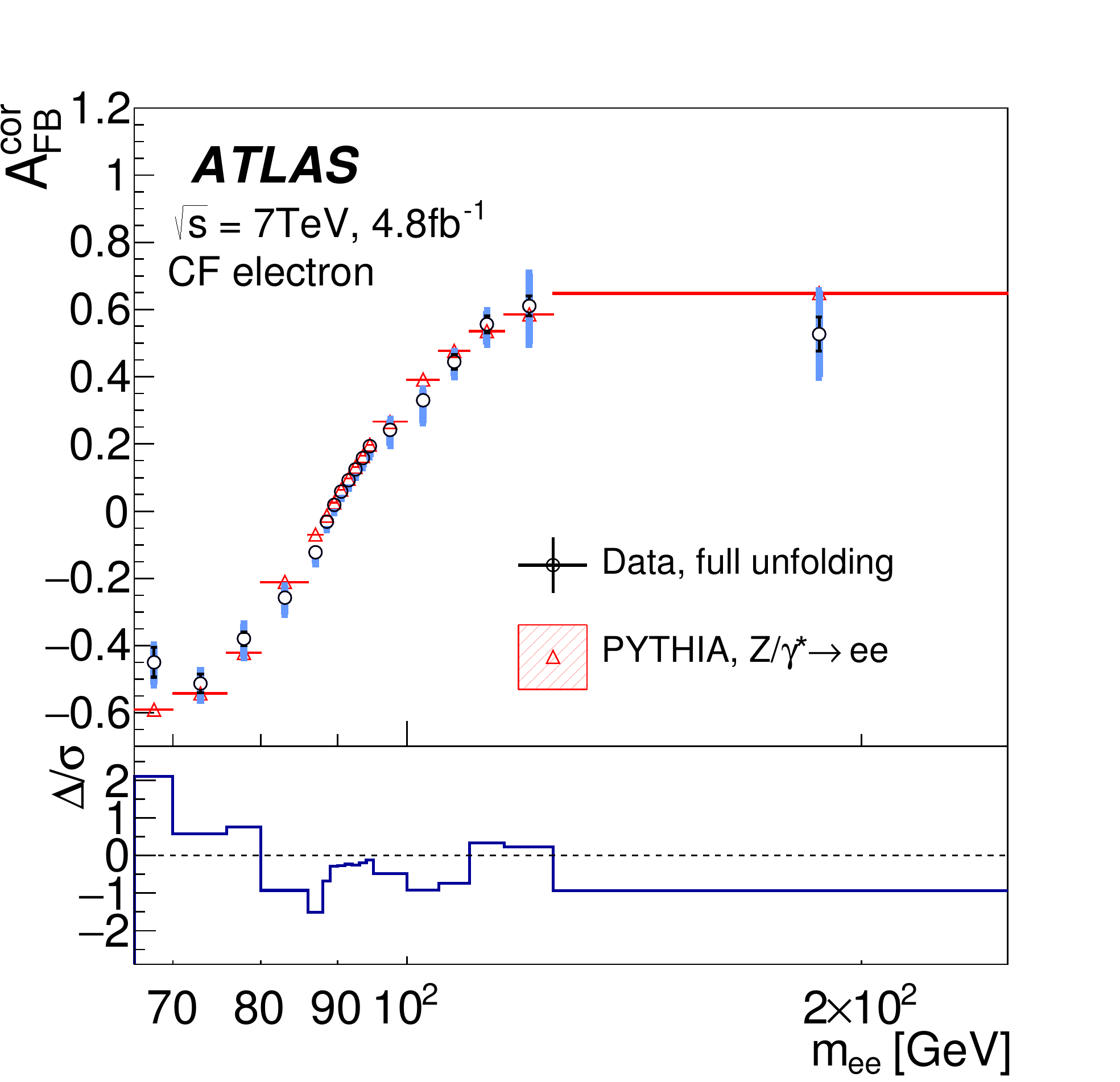} \hspace*{-0.5cm}
\includegraphics[width=5.6cm, height=5.2cm,bb= 0 25 580 420]{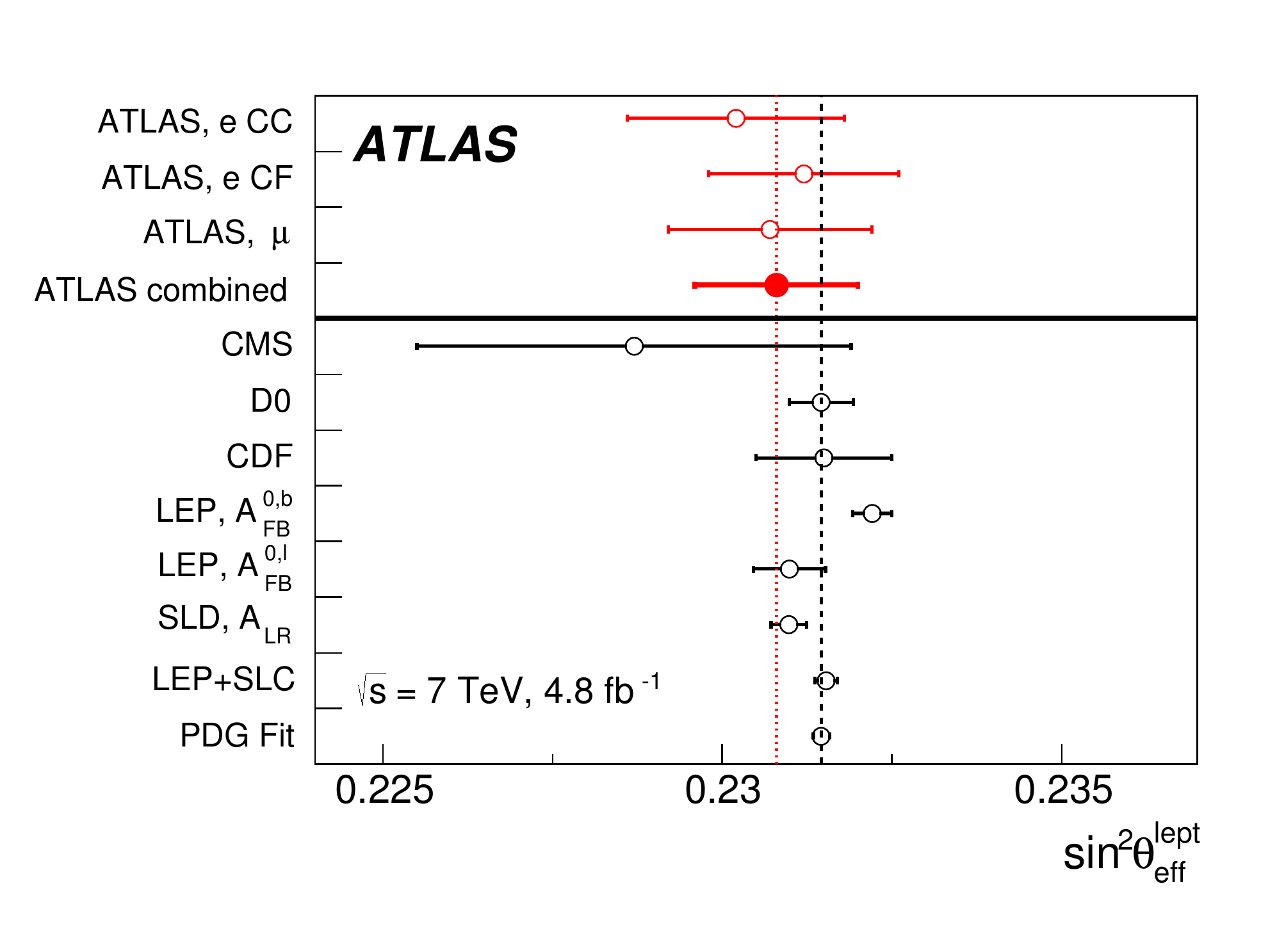} \hspace*{-0.2cm}
\includegraphics[width=5.4cm, height=5.1cm]{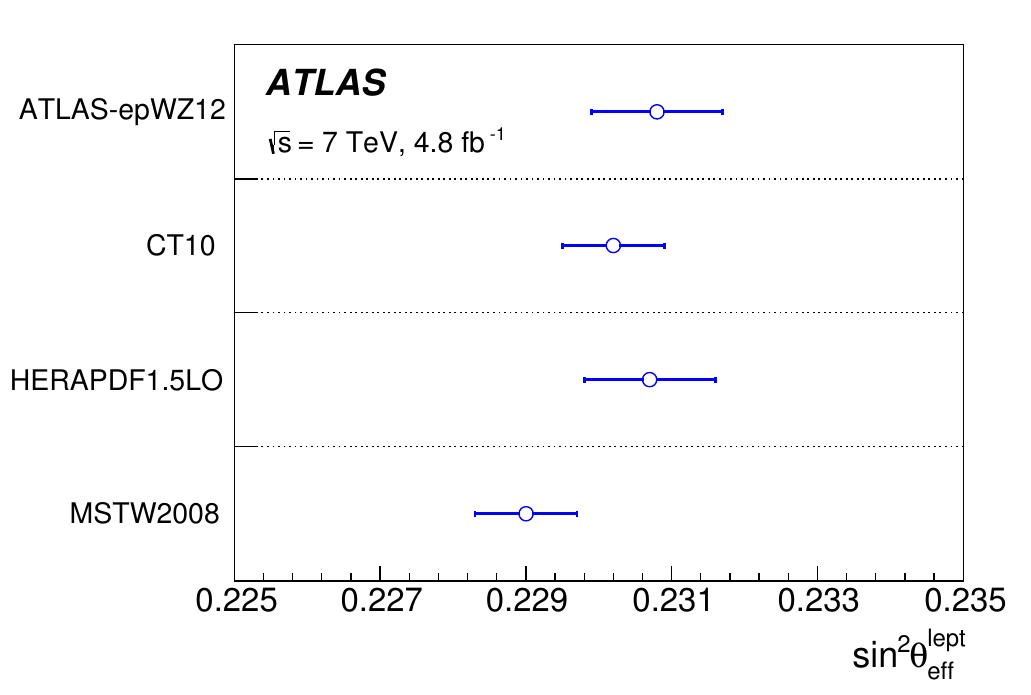} 
\caption{(left) Forward-backward asymmetry values as a function of the di-lepton invariant mass for the central - forward di-electron channel. (middle) Comparison of the measurements of the effective weak mixing angle with other published results. (right) Comparison of the extracted effective weak mixing angle values when using different PDFs in the templates, with only systematic uncertainties shown. 
Taken from Ref.~\cite{ATLAS-Afb}.
\label{fig:Afb}}
\end{figure}

This measurement is also one of the first high precision studies preparing the way towards the ultimate precision measurement at the LHC, the measurement of the W boson mass from the lepton transverse momentum and the lepton - neutrino transverse mass distributions, with a target uncertainty below 10 MeV, dominated by our limited knowledge of the PDFs and the W transverse momentum distribution. Studies of the Drell-Yan process are thus essential to constrain the above QCD systematics.
 
Measurements of W and Z boson production in association with jets test higher order theoretical calculations as well as help us to understand an important background source in searches for BSM physics. W + jets studies~\cite{ATLAS-Wjets} cover topologies with up to 7 jets and jet energies up to 1~TeV. None of the models provide an accurate description of all studied distributions as illustrated in Figure~\ref{fig:Wjets}. The dominant systematic uncertainties come from jet energy scale and at high jet multiplicities from the top pair background.

\begin{figure}[!h]
\centering
\includegraphics[width=8.2cm]{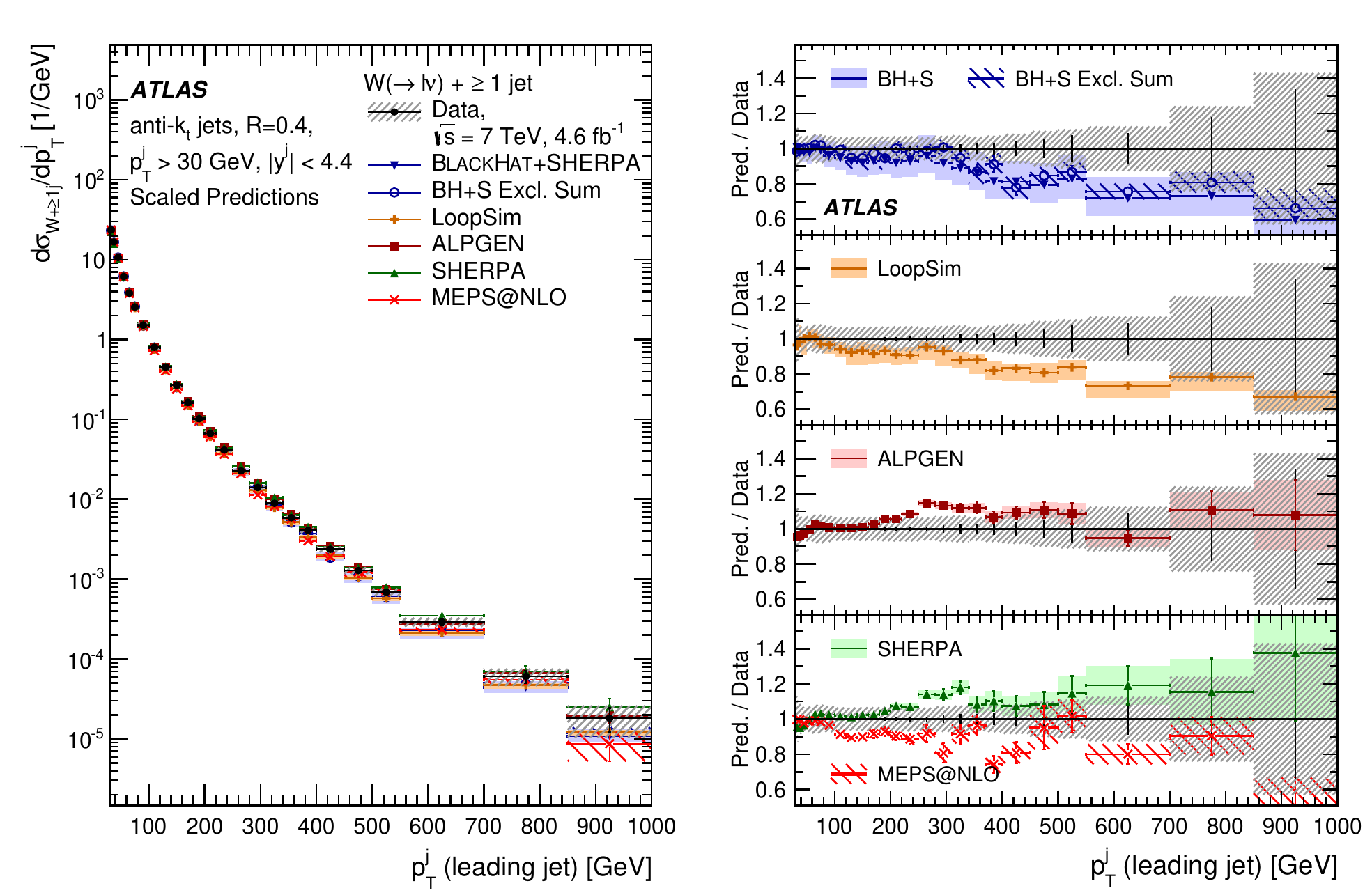} 
\includegraphics[width=8.2cm]{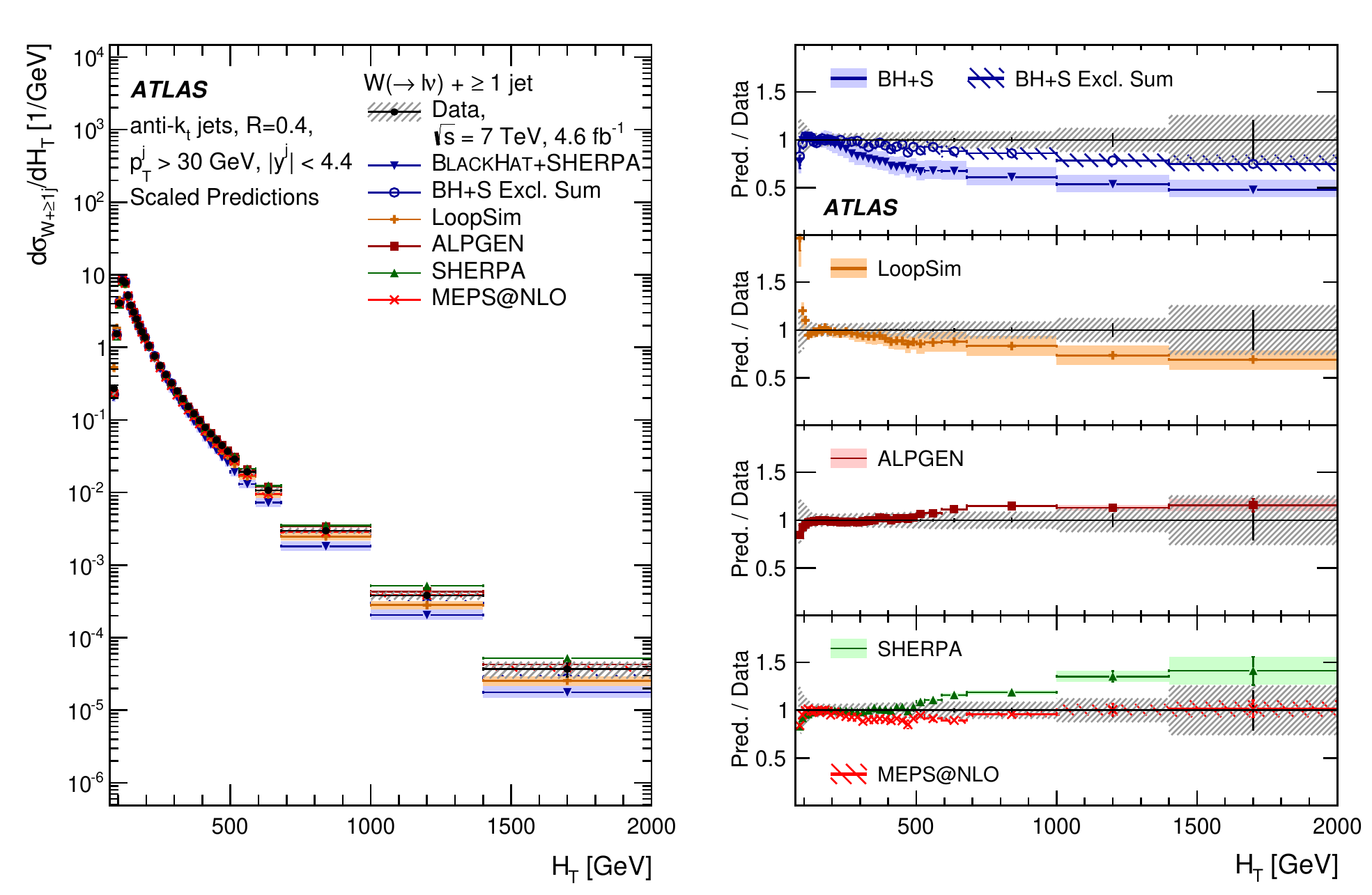} 
\caption{Cross-section for the production of a W boson accompanied by jets as a function (left) of the leading jet transverse momentum and (right) of the $H_\mathrm{T}$, the scalar sum of the transverse momentum of all identified objects in the final state in events with at least one jet. The ratios between the different theoretical calculations and the observed data are also shown.
Taken from Ref.~\cite{ATLAS-Wjets}.
\label{fig:Wjets}}
\end{figure}

Large cancellations of experimental systematics and non-perturbative QCD effects occur when studying cross-section ratios of W boson and Z boson production~\cite{ATLAS-WZjets}. These measurements are well-described in general by next-to-leading order (NLO) perturbative QCD calculations of BlackHat + Sherpa~\cite{BlackHat}, though discrepancies in specific regions are still present as shown in Figure~\ref{fig:WZjets}. 

\begin{figure}[!ht]
\centering
\includegraphics[width=5.4cm]{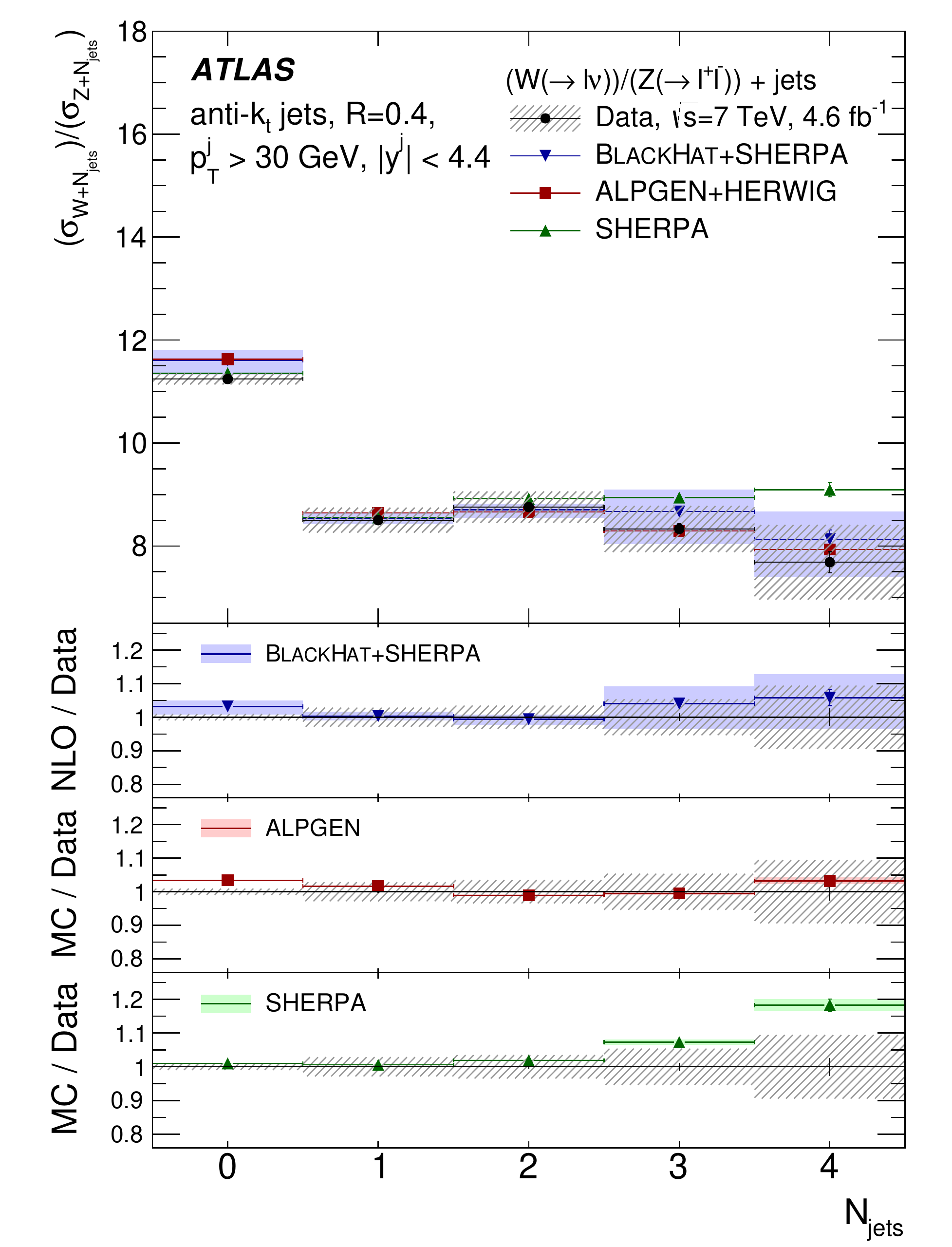} 
\includegraphics[width=5.4cm]{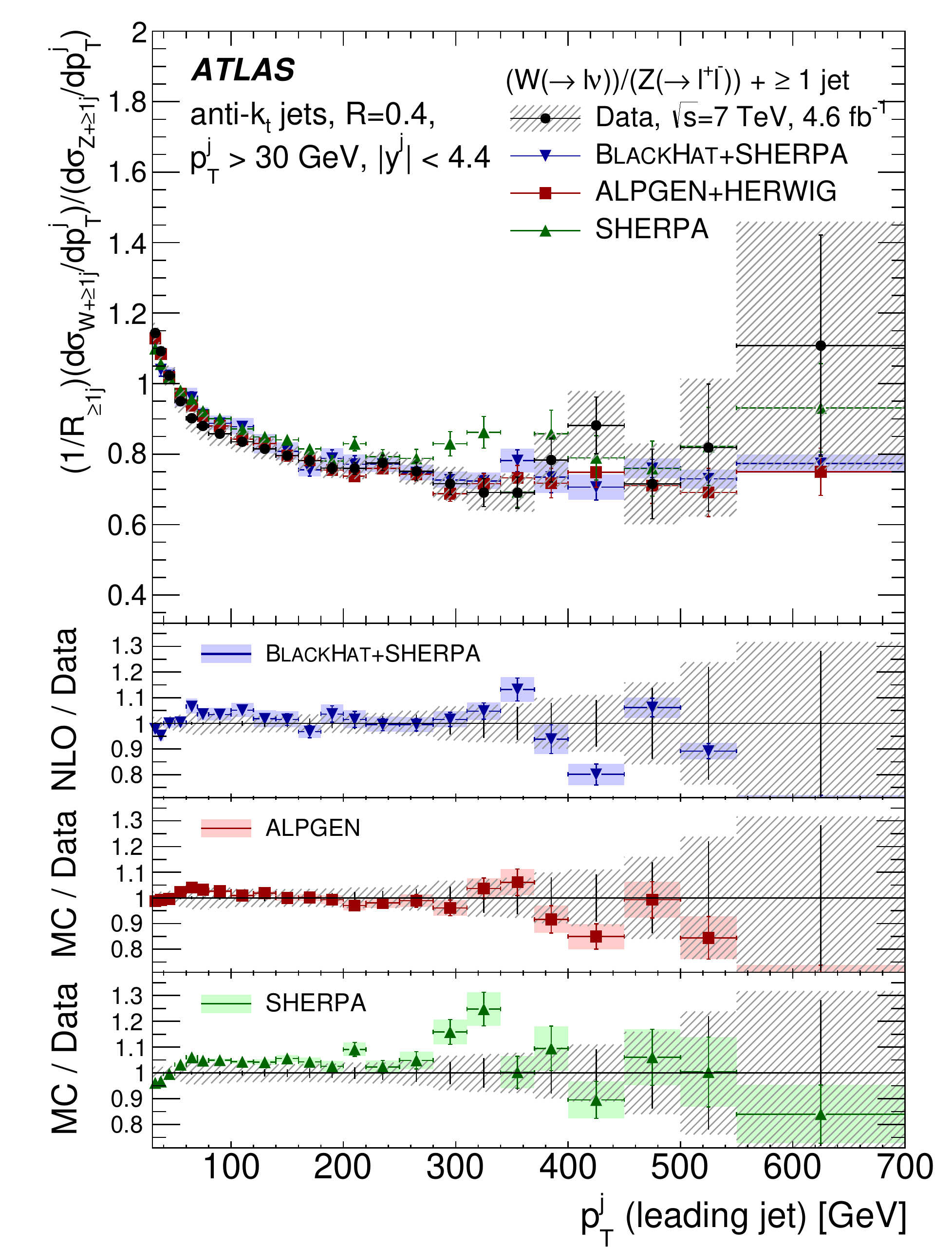} 
\includegraphics[width=5.4cm]{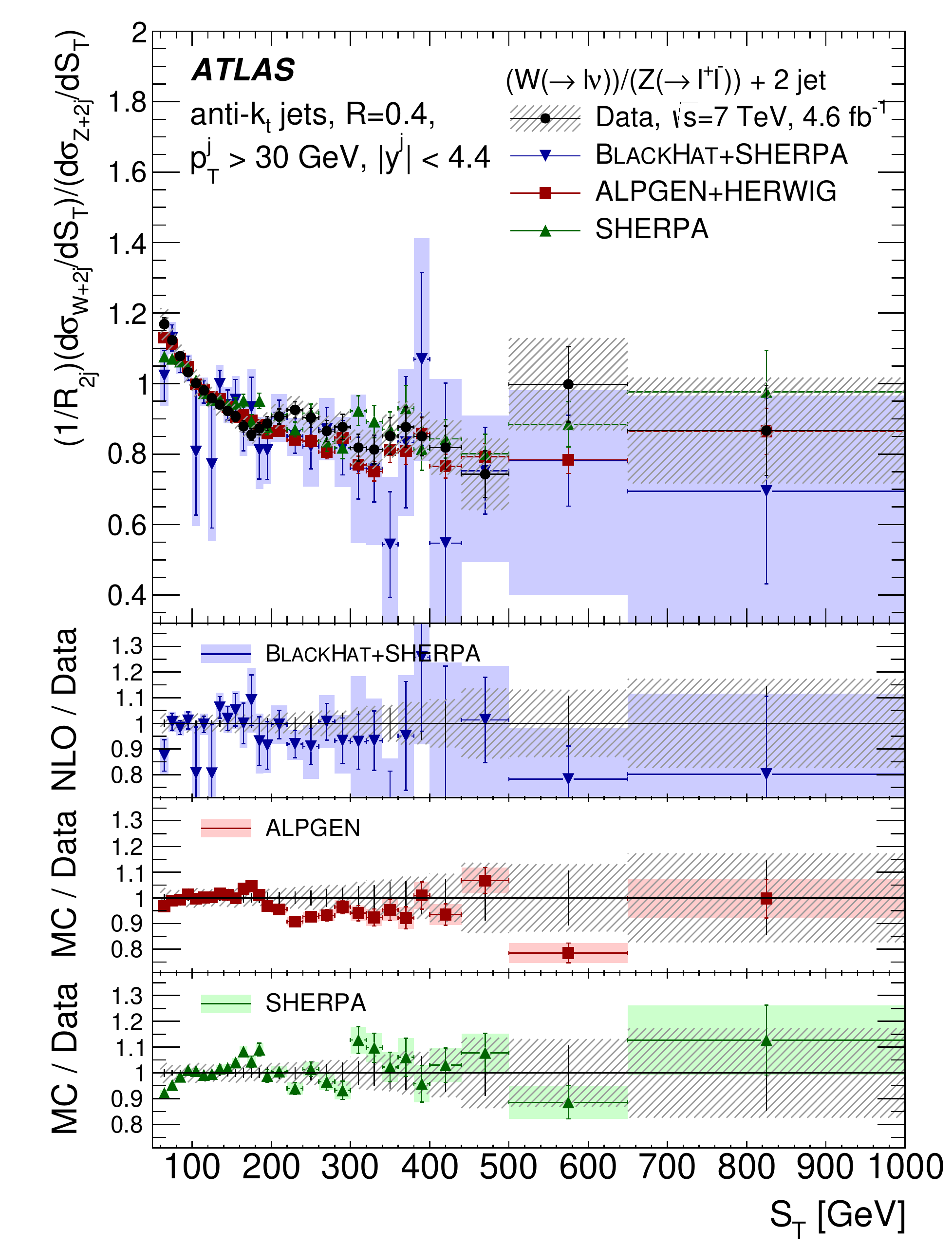} 
\caption{The ratio of W boson and Z boson production cross-sections as a function (left) of the number of jets produced in association with the weak vector boson, (middle) of the leading jet transverse momentum and (right) of the scalar sum of the jet transverse momenta. The ratios between theoretical predictions and the measured values are also shown.
Taken from Ref.~\cite{ATLAS-WZjets}.
\label{fig:WZjets}}
\end{figure}

$Z$+$b$-jet(s) production~\cite{ATLAS-Zbjets} at the LHC is a significant background for Higgs measurements and for New Physics searches. The observed cross-sections are not well described by theoretical calculations as illustrated in Figure~\ref{fig:Zbjets}. In particular, NLO calculations have difficulty with the shape of angular distributions. No preference can be established between four-flavour and five-flavour schemes, as different distributions prefer different calculations. Leading-order (LO) multi-leg generators reproduce the shapes better while underestimating the cross-section itself.

\begin{figure}[!ht]
\centering
\includegraphics[width=5.4cm]{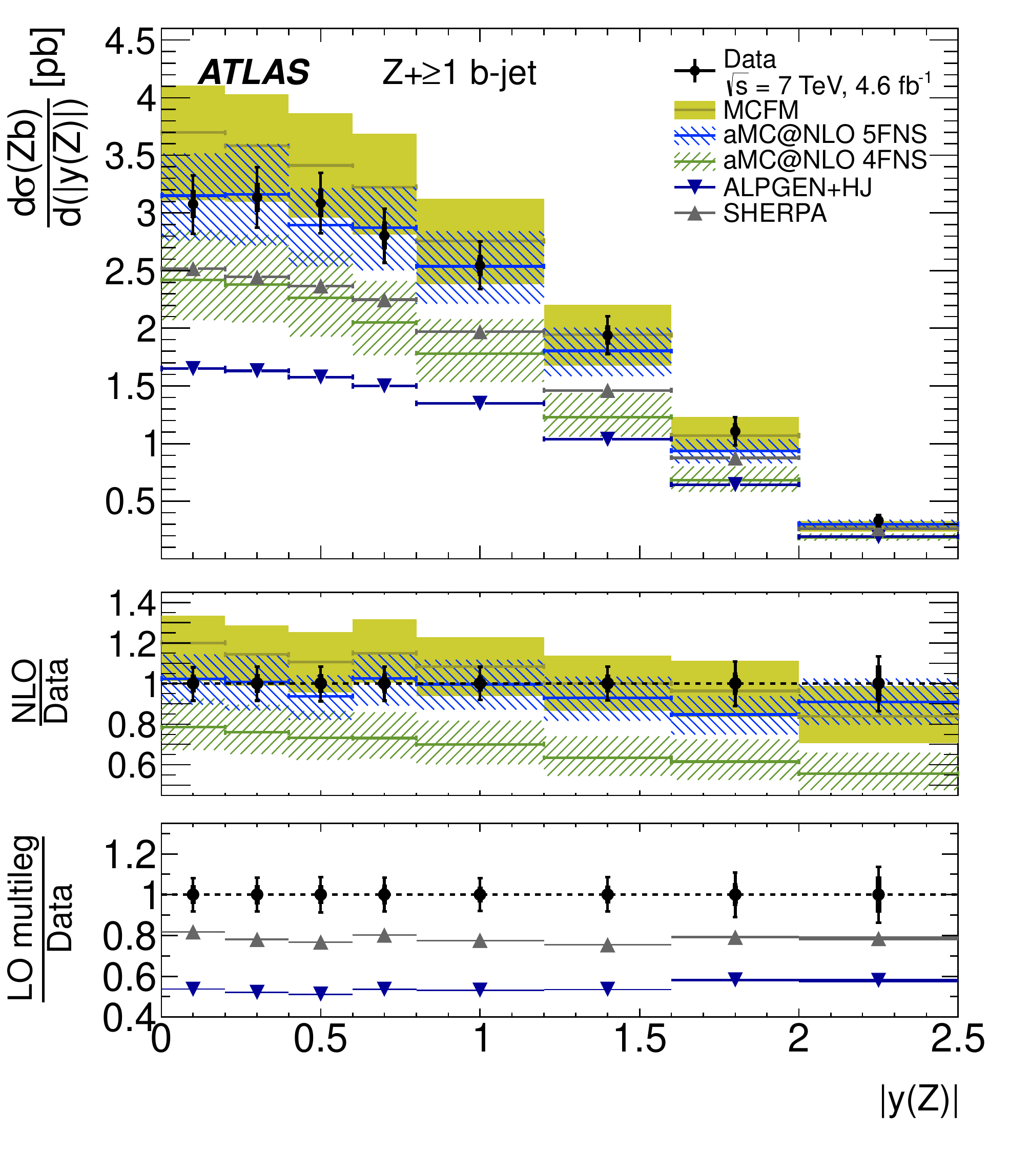} 
\includegraphics[width=5.4cm]{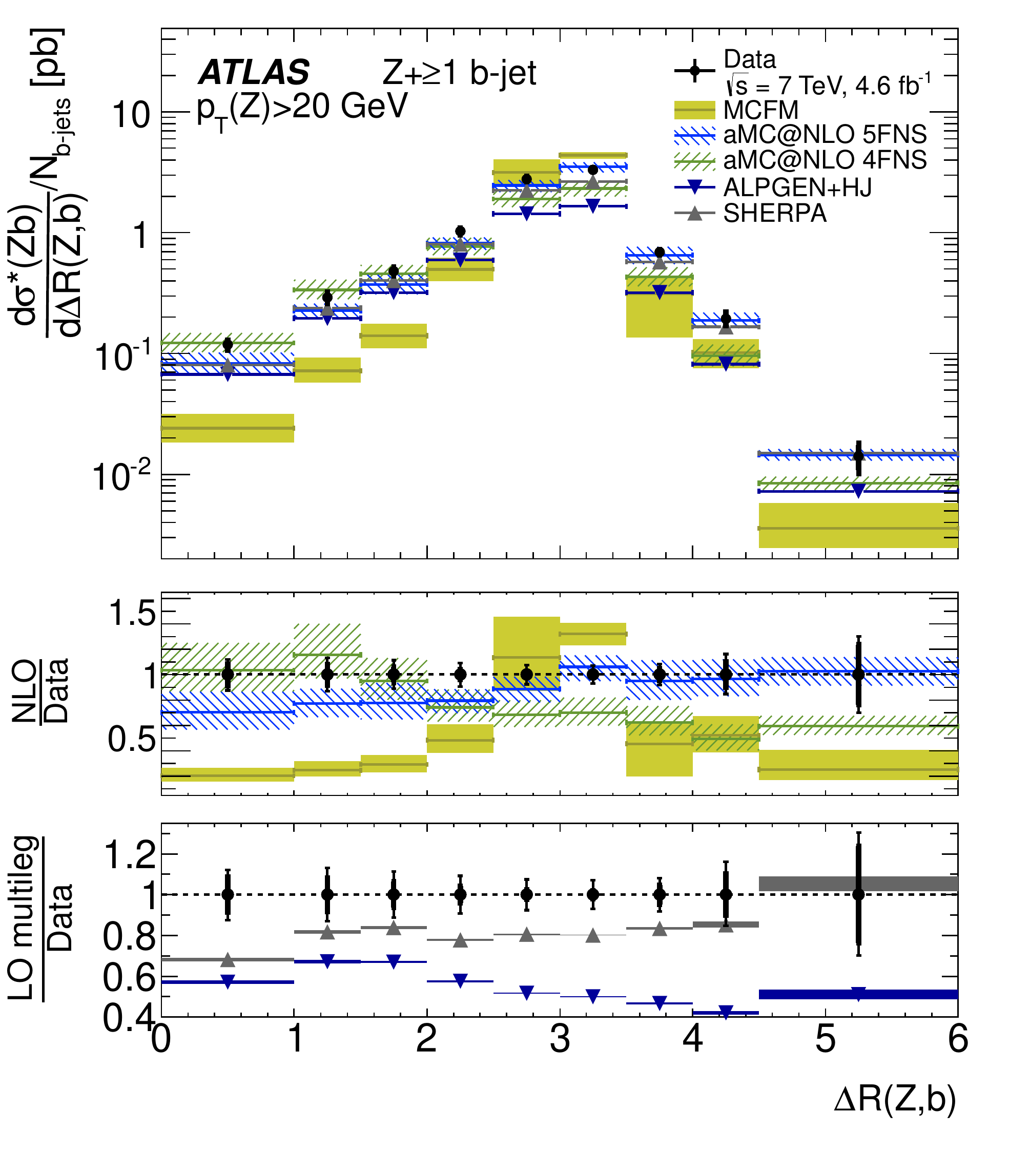} 
\includegraphics[width=5.4cm]{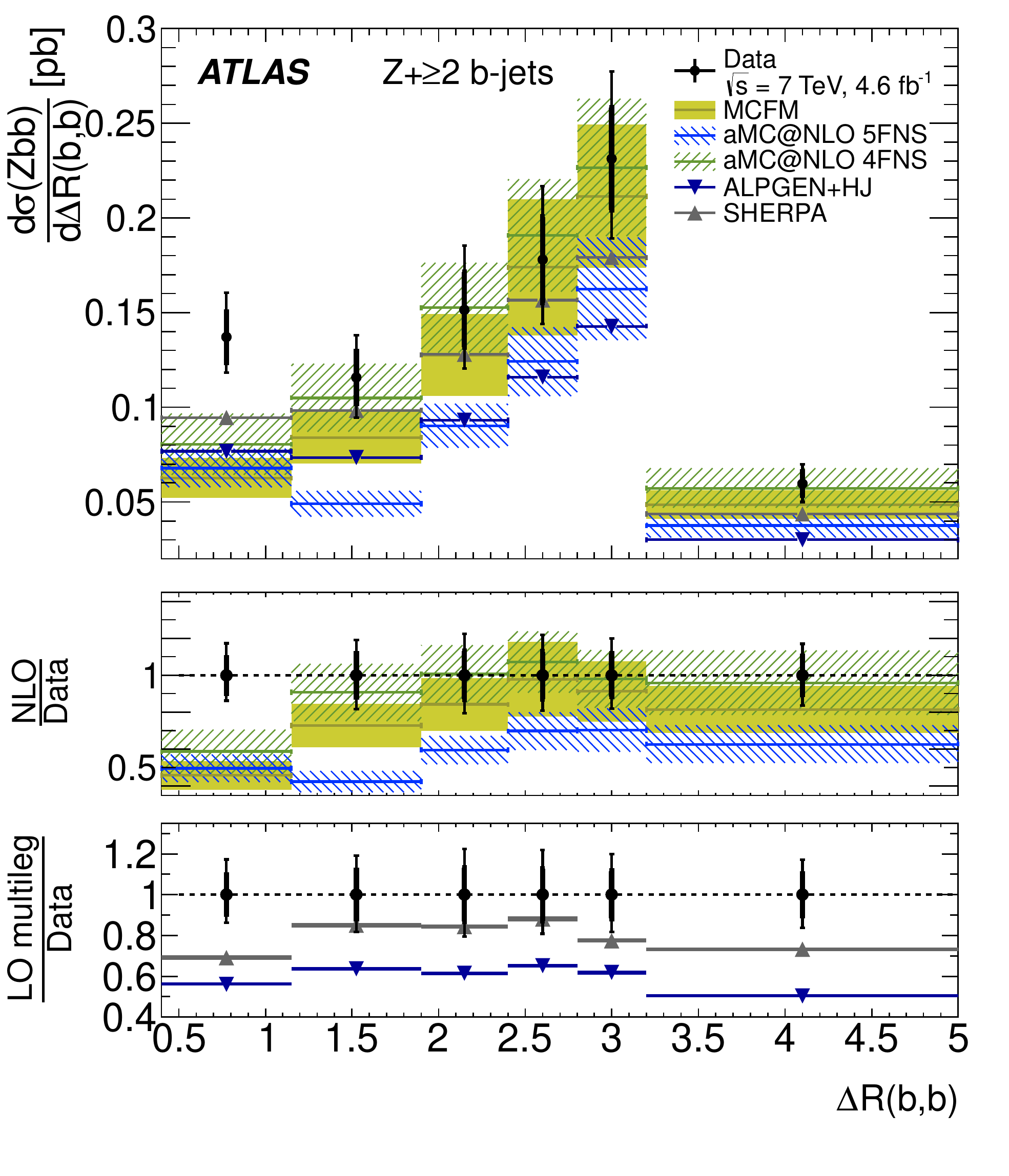} \vspace*{-0.5cm}

\caption{The inclusive b-jet cross-section as a function (left) of the Z boson rapidity (sensitive to the PDFs) and (middle) as the angular distance between the Z boson and the b-jet (sensitive to additional radiation). (right) Cross-section of $Zb\bar{b}$ production as a function of the angular distance between the two b-jets (sensitive to the production mechanism).
Taken from Ref.~\cite{ATLAS-Zbjets}.
\label{fig:Zbjets}}
\end{figure}

\section{Multi-boson production and electroweak physics}

Studying di- and tri-boson final states - and in particular vector boson (V) scattering - plays an important role in probing the electroweak sector of the SM and the presence of New Physics. In the SM, unitarity in VV scattering is restored by Higgs exchange. If the HVV coupling is not exactly the SM value, unitarity is either not realised or delayed until a new high-mass state enters. Thus even if no New Physics is observed directly due to the limited energy reach of LHC or large backgrounds hiding a small signal, VV scattering measurements could reveal its existence. 
 
\begin{figure}[!htbp]
\centering
\includegraphics[width=16cm,height=10cm]{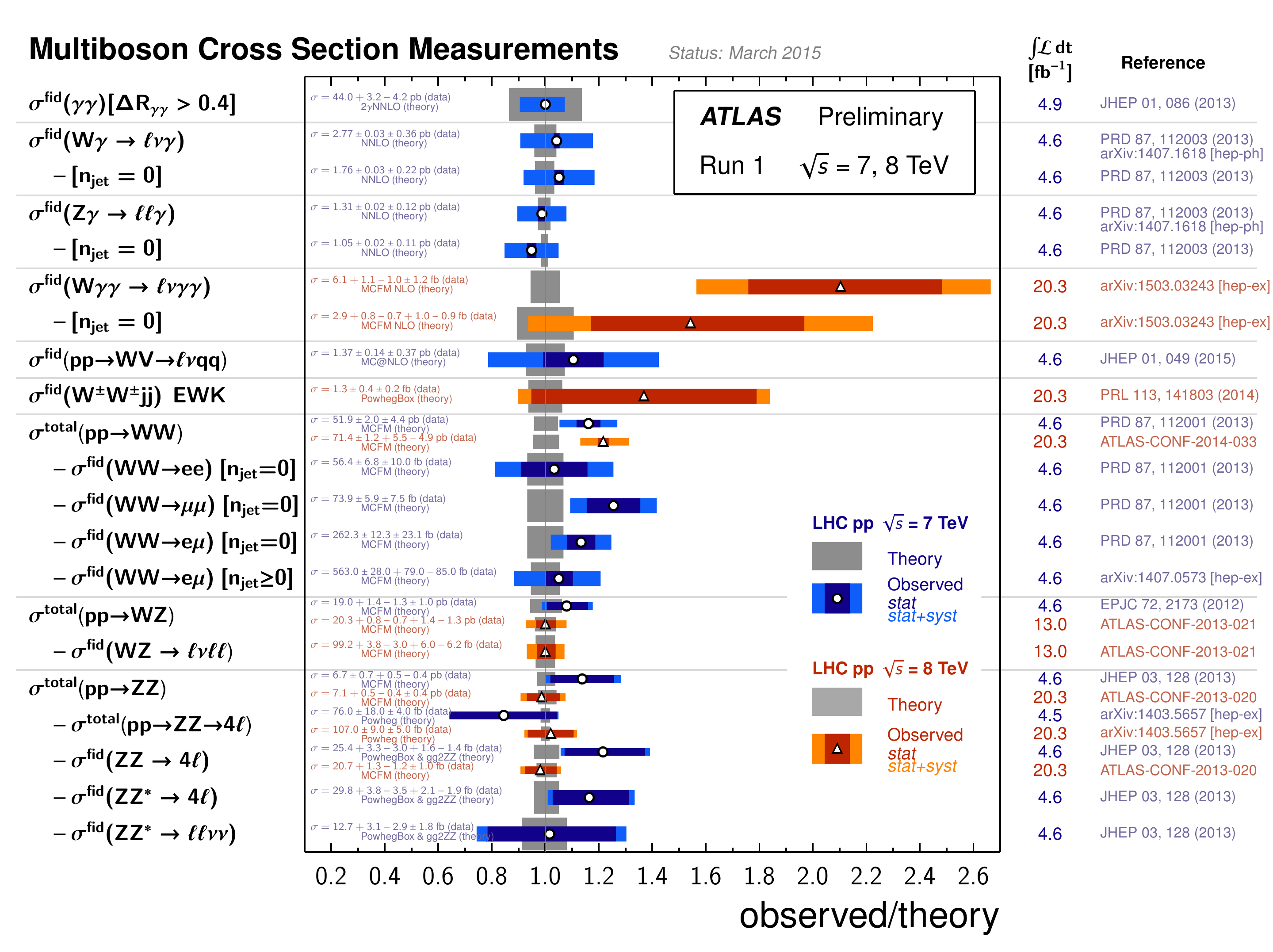}
\caption{The data/theory ratio for several di-boson total and fiducial production cross-section measurements, corrected for leptonic branching fractions. All theoretical expectations were calculated at NLO or higher. 
Taken from Ref.~\cite{ATLAS-SMSummary}.
\label{fig:VVSummary}}
\end{figure}

Multi-boson production cross-section measurements, summarised in Figure~\ref{fig:VVSummary}~\cite{ATLAS-SMSummary}, show in general a good agreement with theoretical predictions. The most discrepant value is that of the WW cross-section where a 2.1$\sigma$ excess is seen in data with respect to the MCFM~\cite{MCFM} calculation at NLO in QCD including off-shell bosons and decays. Investigations point toward underestimated uncertainties related to missing higher order contributions as well as to the jet veto (necessary to suppress the large top quark pair background). Detailed measurements of the differential cross-sections help to uncover these effects by providing information on event kinematics, production modes, and the presence of additional radiation. 

The high-energy tails of the measured distributions are especially sensitive to New Physics and can be used to constrain anomalous gauge couplings. In Figure~\ref{fig:Wx} the di-jet transverse momentum $p_{\mathrm{T}jj}$ is shown for a $WW / WZ \rightarrow \ell\nu jj$ selection~\cite{ATLAS-VW}. The lower panel illustrates how the presence of an anomalous triple gauge coupling (aTGC)~\cite{TGCdef} would lead to an excess of events at high $p_{\mathrm{T}jj}$. The obtained limits are shown in Figure~\ref{fig:AnomCoup} together with constraints from other ATLAS measurements and from other experiments.

\begin{figure}[!ht]
\centering
\includegraphics[width=5.4cm]{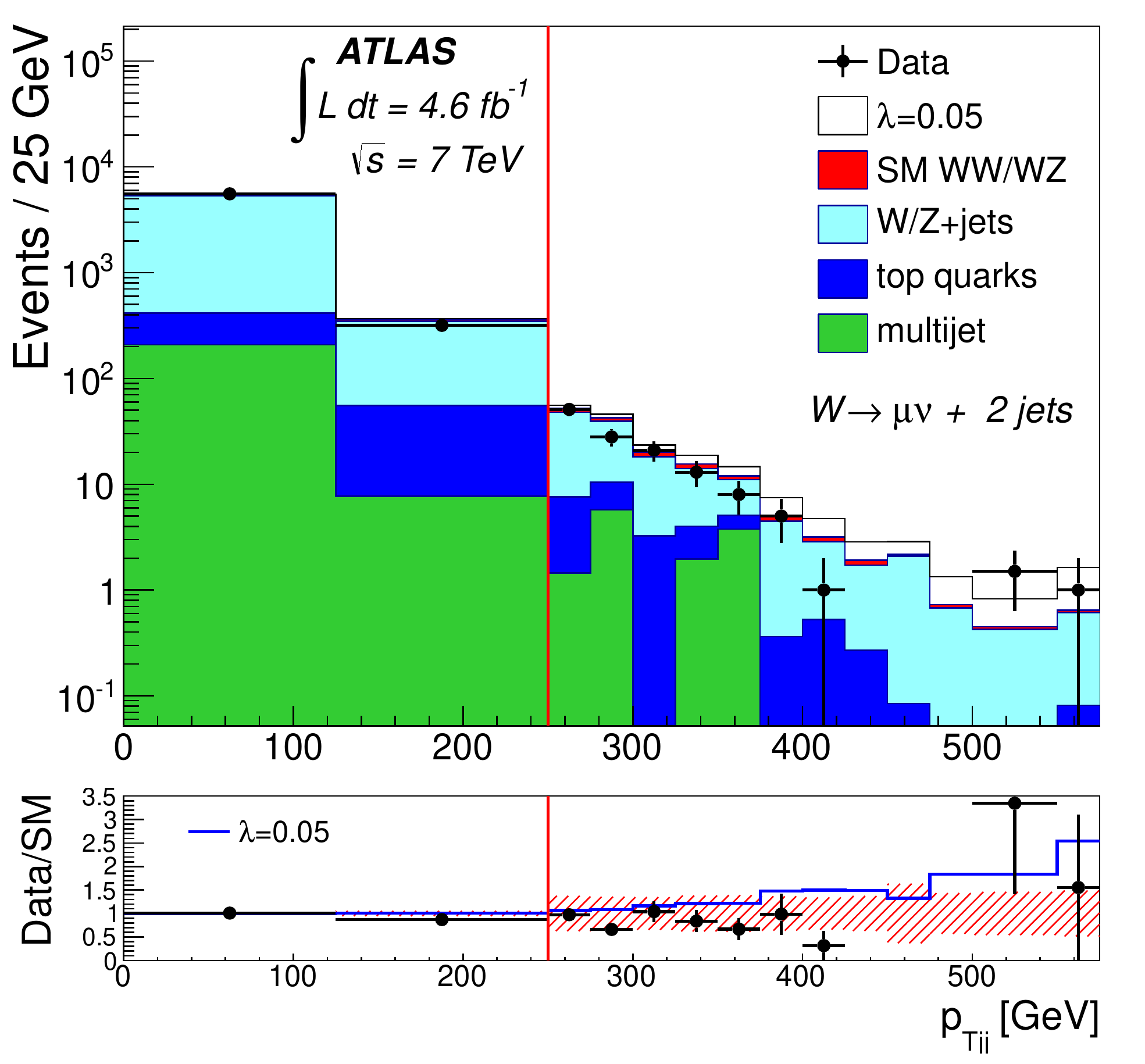} 
\includegraphics[width=5.4cm]{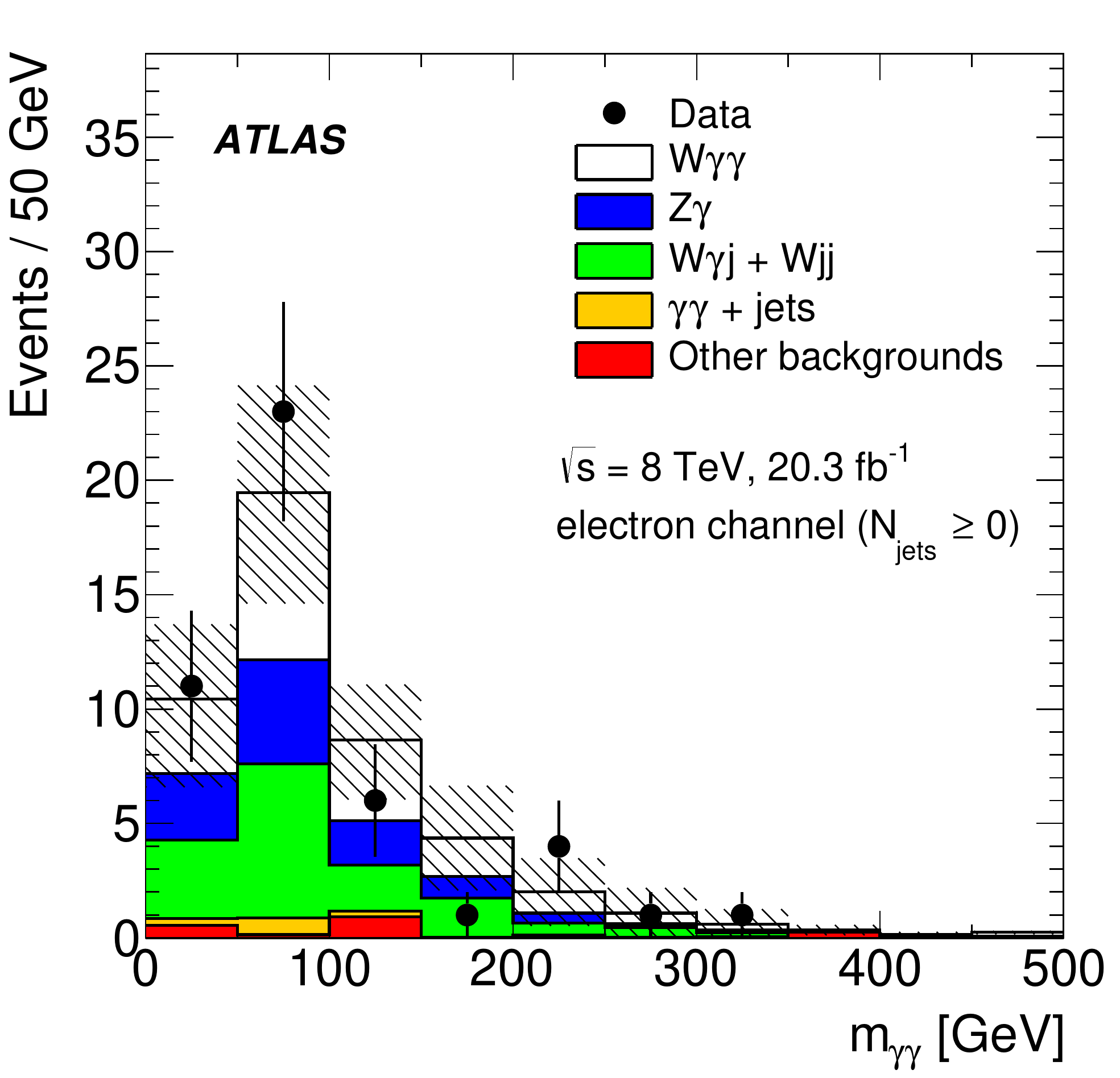} 
\includegraphics[width=5.4cm]{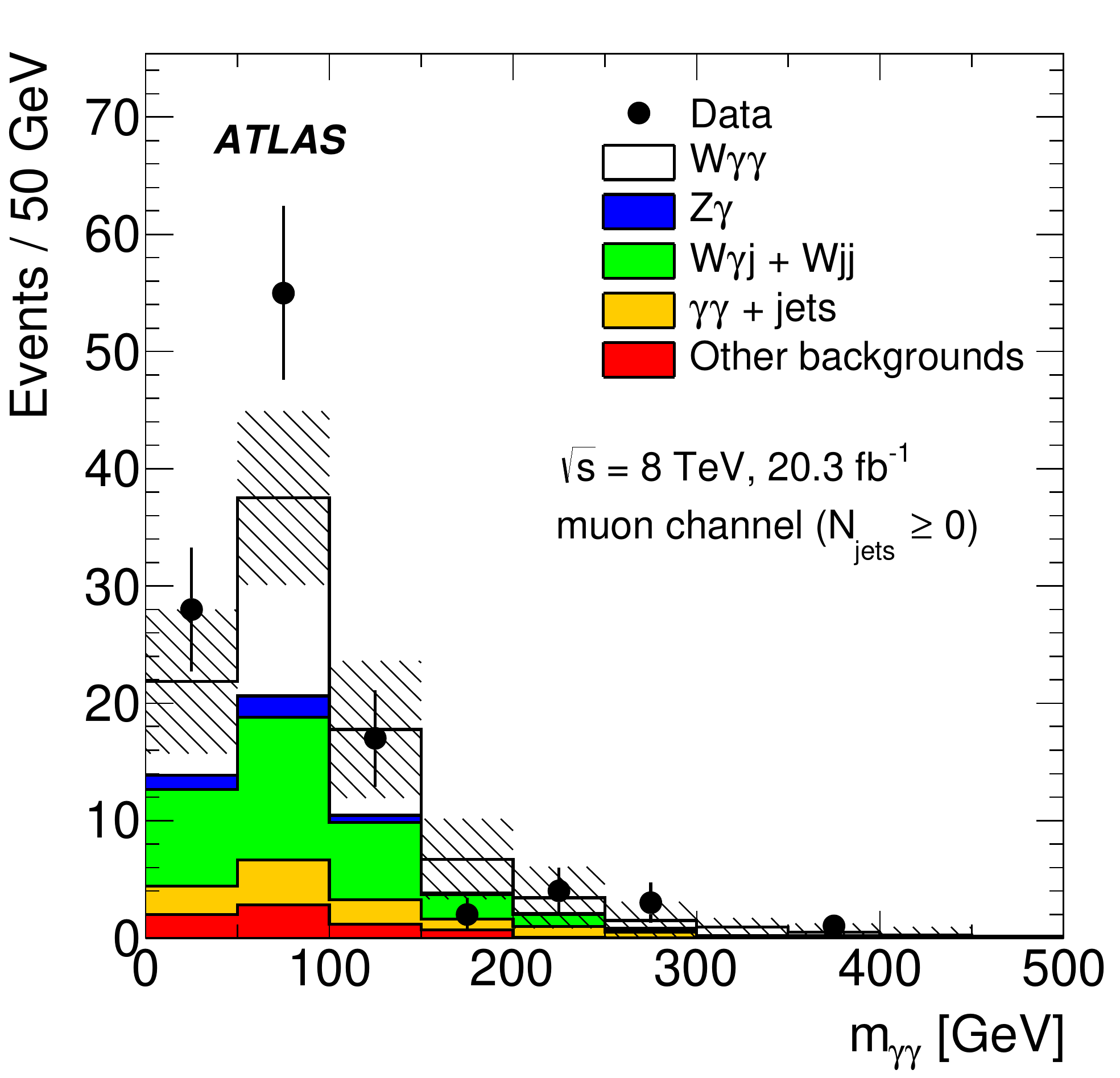} 
\caption{ (left) The observed distribution of the transverse momentum of the two jets, compared to the expectation for SM signal plus background in the muon channel of the $WW/WZ \rightarrow \ell\nu jj$ analysis. The effect of an aTGC is shown for comparison on top of the SM predictions.  (middle and right)  Di-photon invariant mass distribution in the electron and muon channels in the $W\gamma\gamma \rightarrow \ell\nu\gamma\gamma$ analysis.
Taken from Refs.~\cite{ATLAS-VW,ATLAS-Wgg}.
\label{fig:Wx}}
\end{figure}

\begin{figure}[!hbt]
\centering
\includegraphics[width=8.1cm,height=8.2cm, bb=0 0 550 700]{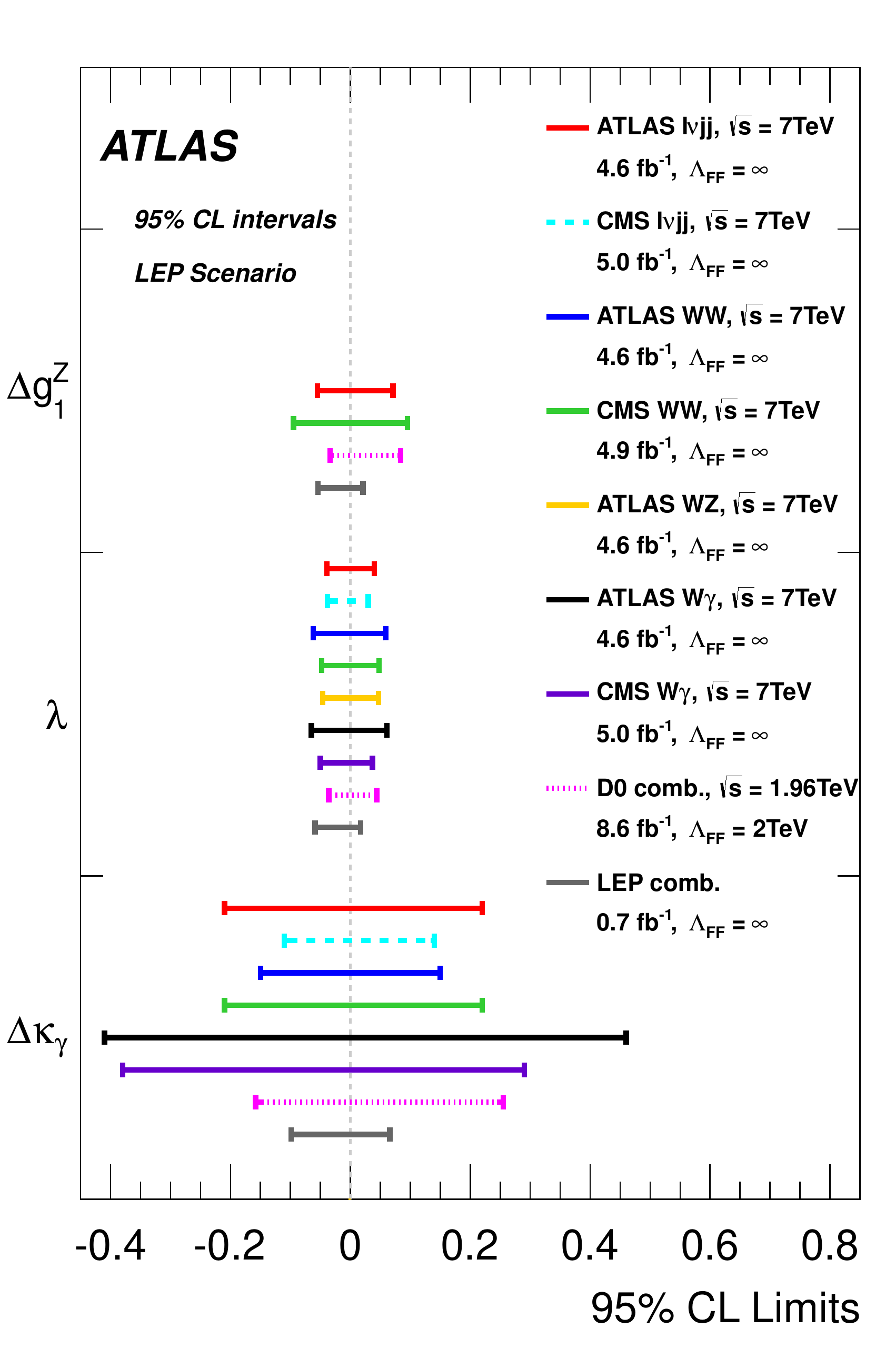} 
\includegraphics[width=8.1cm,height=7cm, bb=0 -20 570 500]{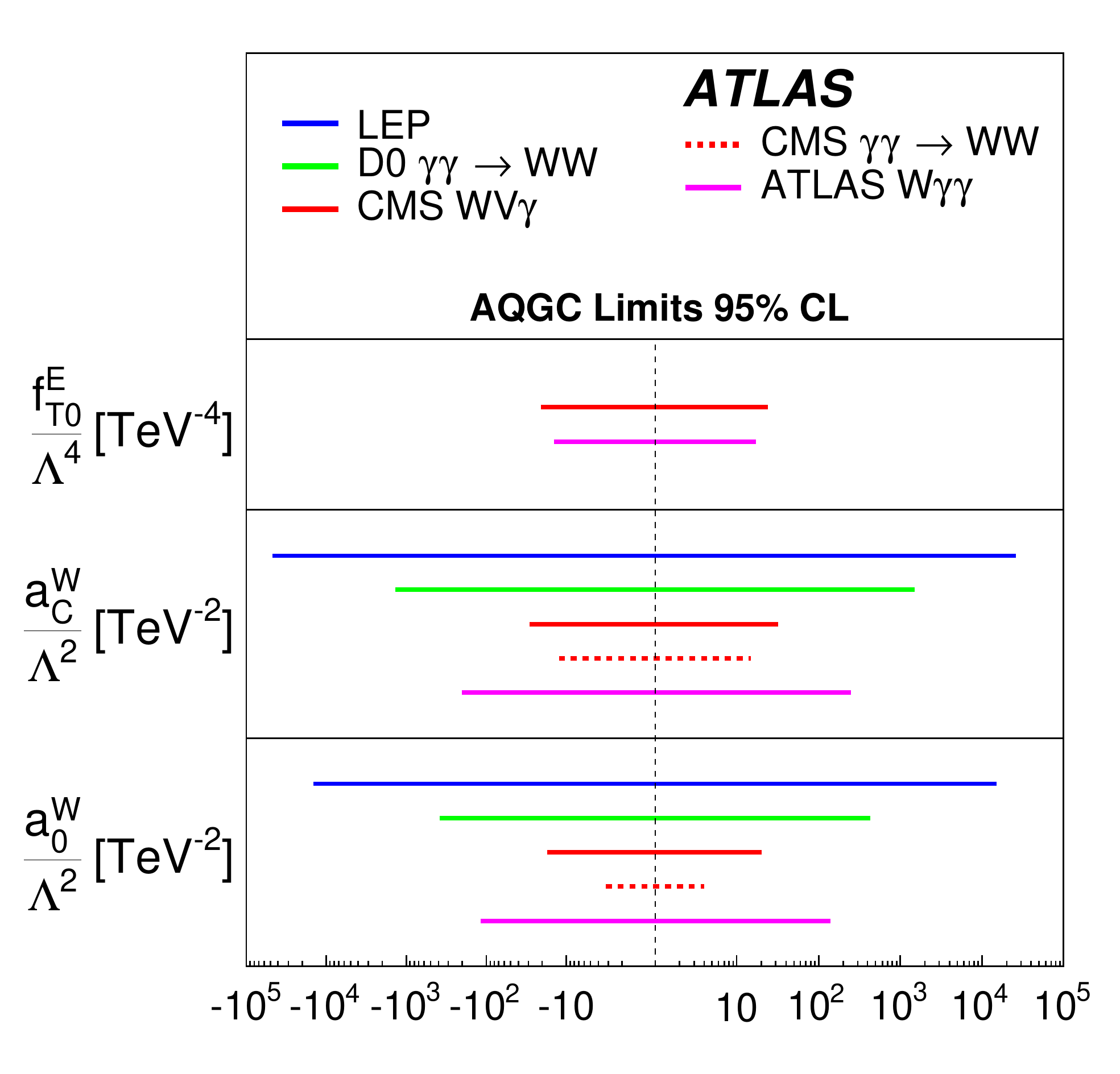} 
\caption{Summary of limits on anomalous (left) triple and (right) quartic gauge coupling parameters.
Taken from Refs.~\cite{ATLAS-VW,ATLAS-Wgg}.
\label{fig:AnomCoup}}
\end{figure}

The first evidence for triple gauge boson production exceeding the significance of 3$\sigma$ comes from the study of $W\gamma\gamma$ production~\cite{ATLAS-Wgg}. The measured inclusive fiducial cross-section shows a 1.9$\sigma$ excess over the NLO prediction of MCFM. The measured exclusive cross-section requiring the absence of jets, while still above the expectation, agrees within systematics with MCFM. The high-mass region of the di-photon mass distribution shown in Figure~\ref{fig:Wx} can be used to constrain anomalous quartic gauge couplings (aQGC)~\cite{QGC-Wgg-def}. The achieved constraints are compared to other measurements in Figure~\ref{fig:AnomCoup}.

WW scattering can be studied in the $W^\pm W^\pm jj$ final state requiring two forward jets and little hadronic activity between them. Figure~\ref{fig:WWjj} shows the observed distribution of the rapidity difference between the two jets $|\Delta y_{jj}|$ and illustrates how EW production inhabits the high $|\Delta y_{jj}|$ region. This rare process is the most sensitive at the LHC to access the VBS region. The ATLAS results~\cite{ATLAS-WWjj} give the first evidence for inclusive and electroweak $W^\pm W^\pm jj$ production with 4.5$\sigma$ and 3.6$\sigma$ significance. The measured cross-sections are in agreement with the SM within the large, statistics dominated uncertainties. The measurement can also be used to constrain the aQGC~\cite{QGC-WWjj-def} values as shown in Figure~\ref{fig:WWjj}.

So far all anomalous coupling measurements are compatible with the SM and no hint of New Physics is visible.

\begin{figure}[!ht]
\centering
\includegraphics[width=5.4cm,height=4cm]{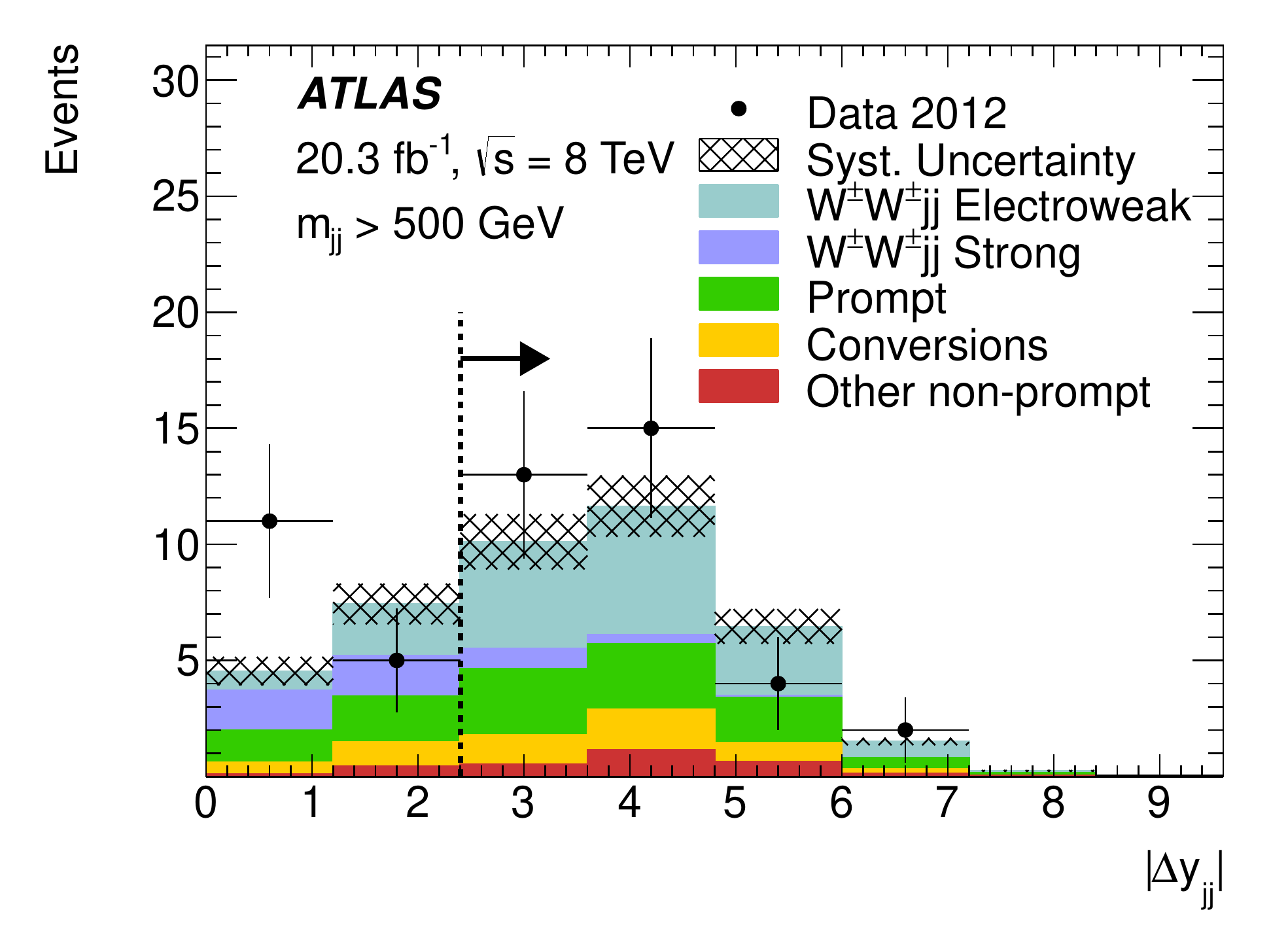} \hspace*{0.05cm}
\includegraphics[width=5.4cm,height=4.2cm]{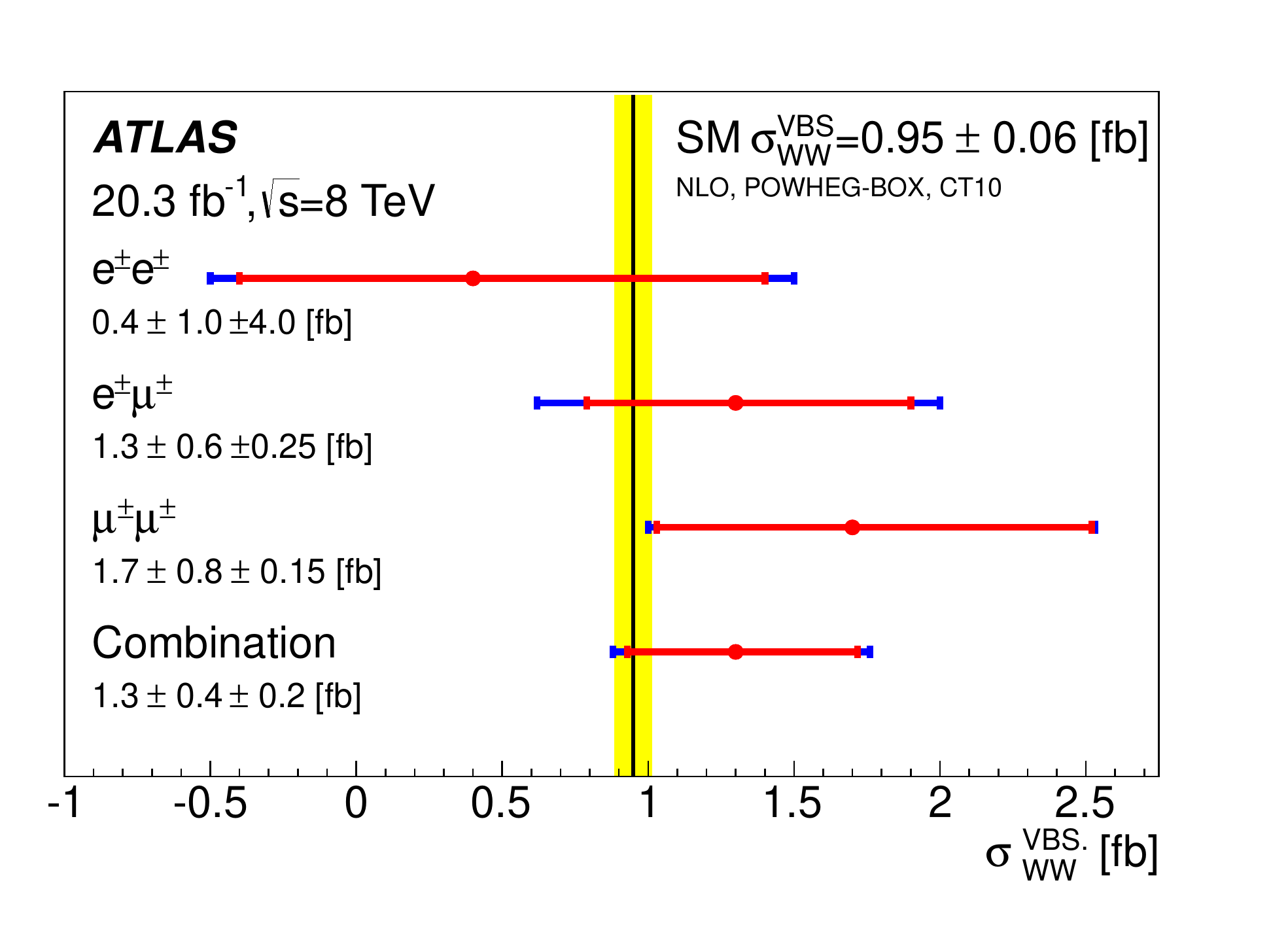} 
\includegraphics[width=5.4cm,height=4cm]{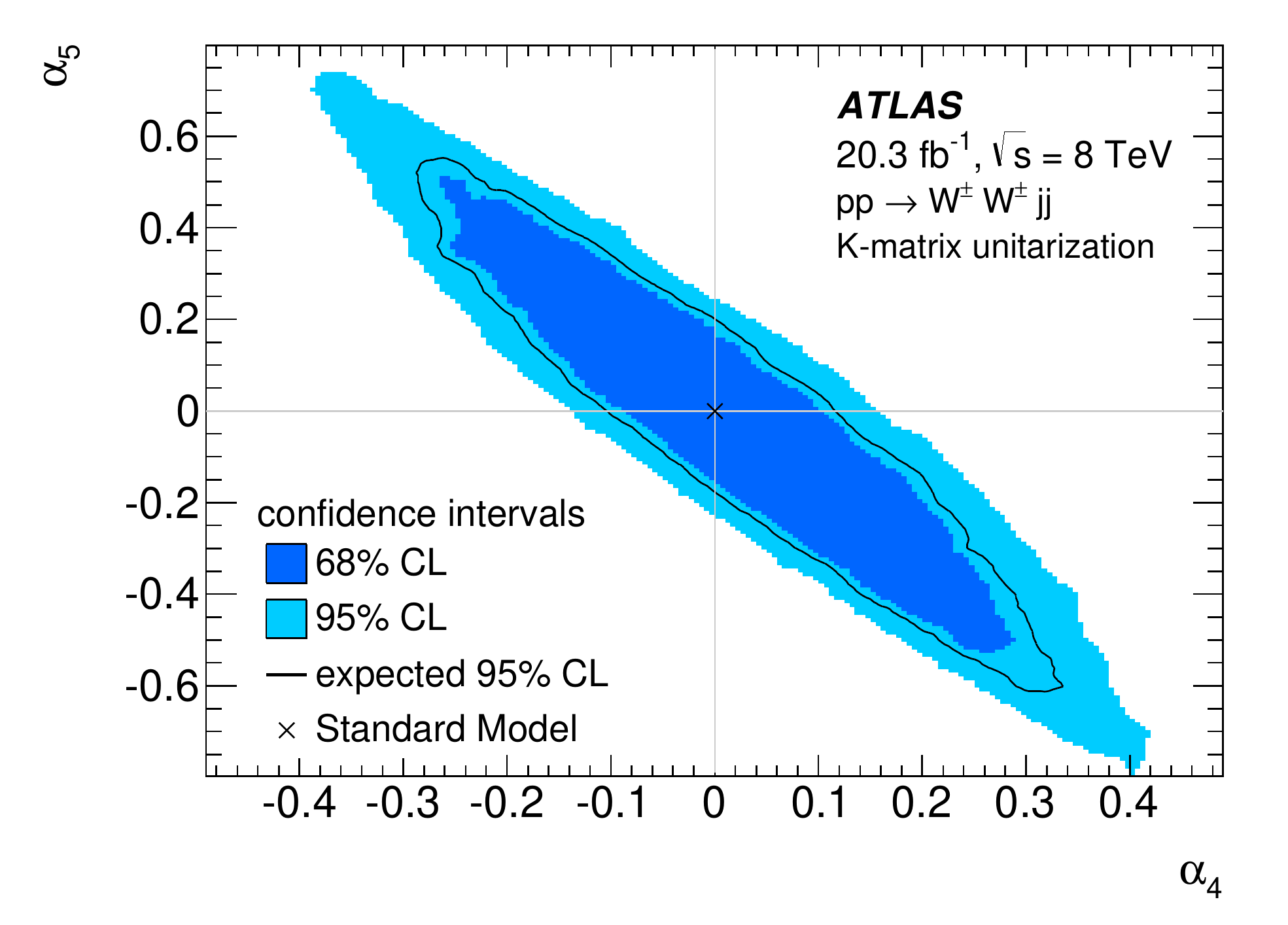} 
\caption{ (left) Distribution of the rapidity difference between two jets in the $W^\pm W^\pm jj$ analysis. (middle) The measured cross-sections for the VBS region compared to the SM prediction. (right) Limits on the anomalous quartic gauge couplings  $(\alpha_4, \alpha_5)$.
Taken from Ref.~\cite{ATLAS-WWjj}.
\label{fig:WWjj}}
\end{figure}

\section{Top quark physics}

The top quark being the heaviest known elementary particle with a mass of about 173~GeV and a Yukawa coupling to the Higgs boson of O(1) is particularly interesting. It is expected to be connected to electroweak symmetry breaking. In BSM models, such as technicolor or other scenarios with strongly coupled Higgs sector, the top couplings can be modified.


The top quark pair-production cross-section is measured in many final states. The total cross-section is well described by next-to-next-to-leading order (NNLO) plus next-to-next-to-leading log (NNLL) theoretical calculations~\cite{ATLAS-TopSummary} as shown in Figure~\ref{fig:topX}. Differential cross-section measurements are performed to understand $t\bar{t}$ production which is the main background to many New Physics searches. These then benefit from the improved modeling uncertainties. Fiducial particle-level differential cross-sections allow for more precise comparison with theoretical calculations than the parton-level ones as illustrated in Figure~\ref{fig:topX} where the transverse momentum dependence is shown. These results~\cite{ATLAS-TopDiffXsec} cover top quark production up to 1.2~TeV transverse momentum using semi-leptonic final states where the hadronically decaying top quark is reconstructed as an anti-kt jet with radius parameter $R = 1$ and identified with jet substructure techniques. These techniques, validated in top quark studies, are also essential in searches for new heavy particles decaying hadronically.

\begin{figure}[!ht]
\centering
\includegraphics[width=5.4cm,height=4.1cm,bb=0 10 570 390]{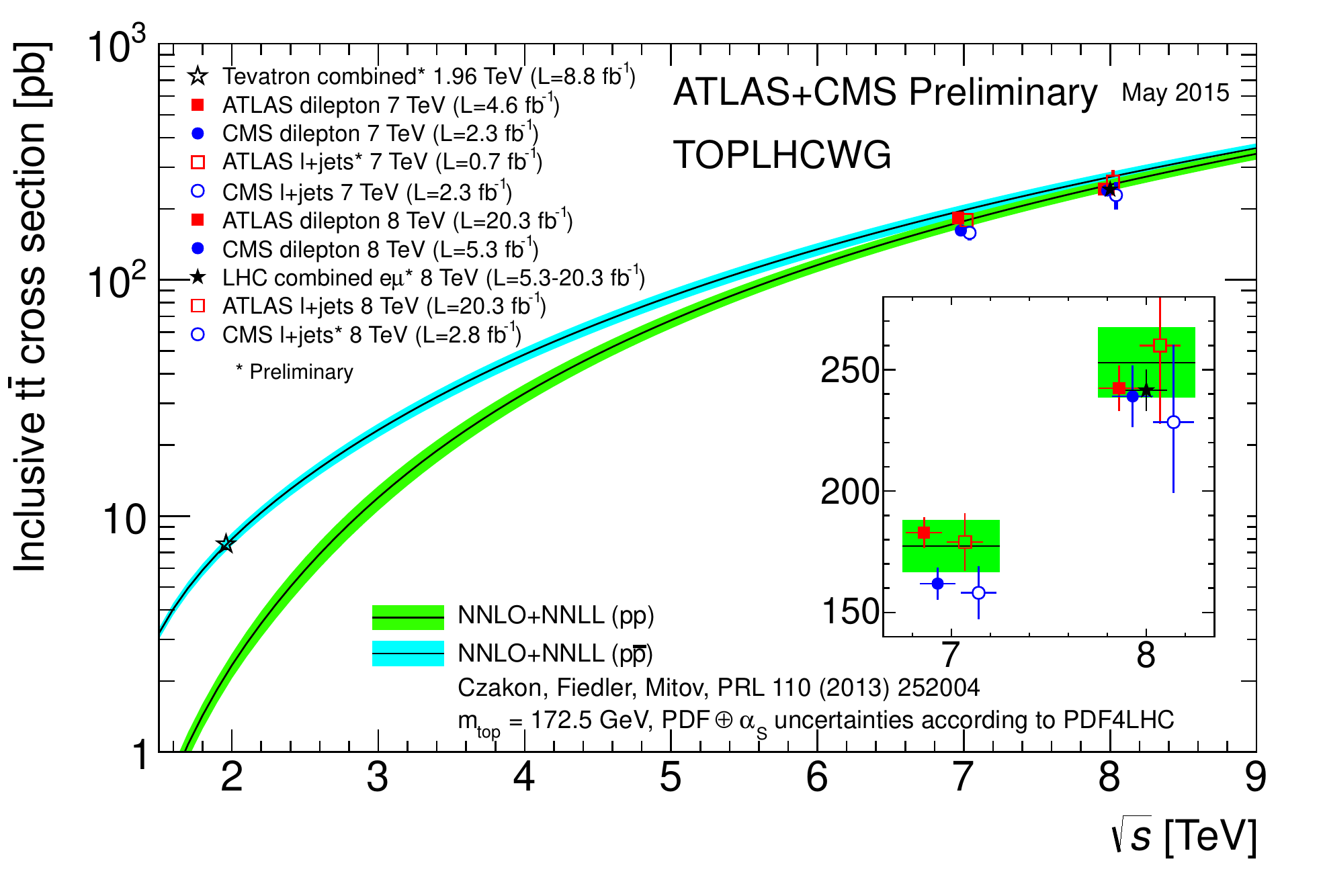} 
\includegraphics[width=5.4cm,height=4cm]{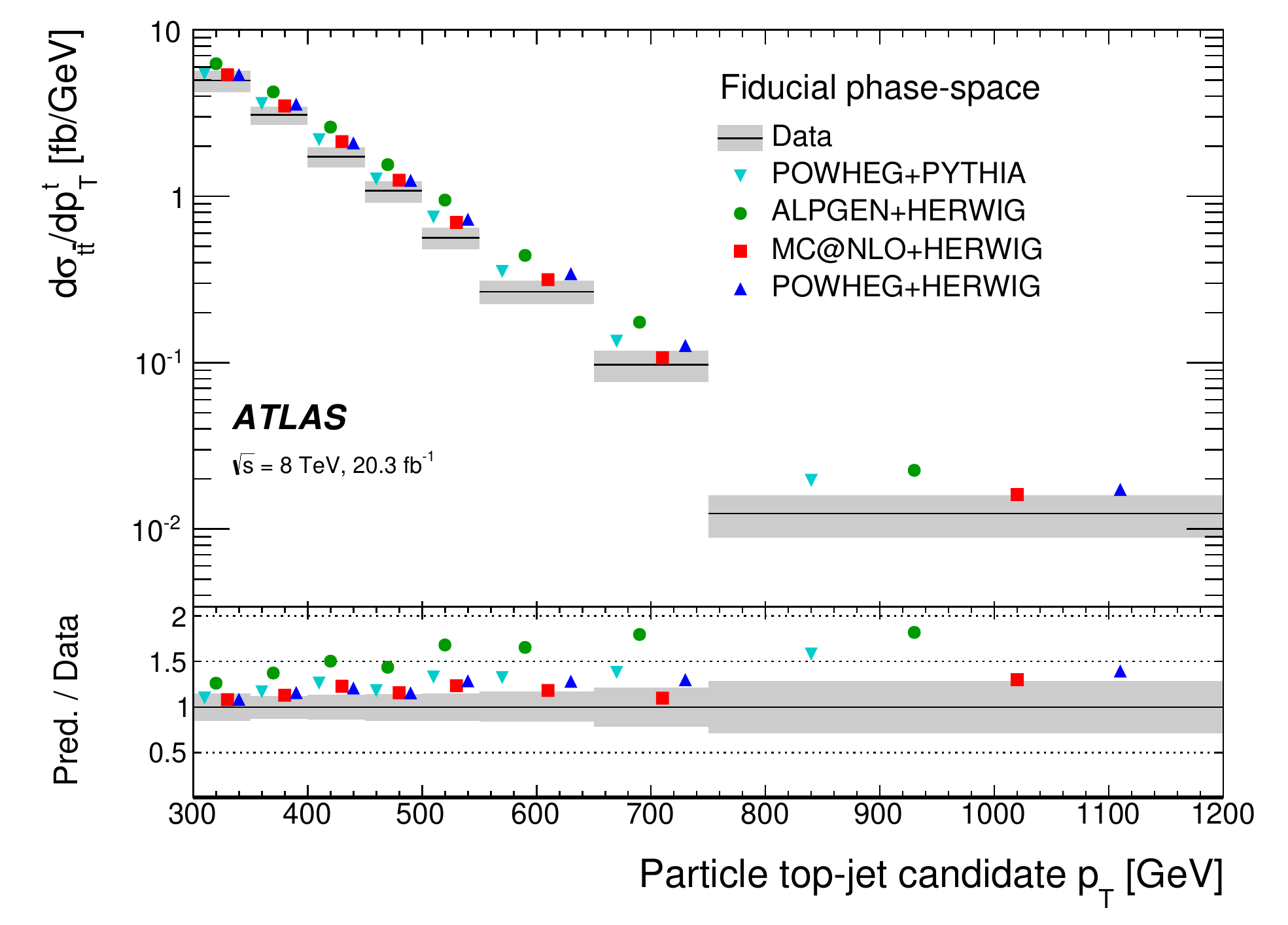}
\includegraphics[width=5.4cm,height=4cm]{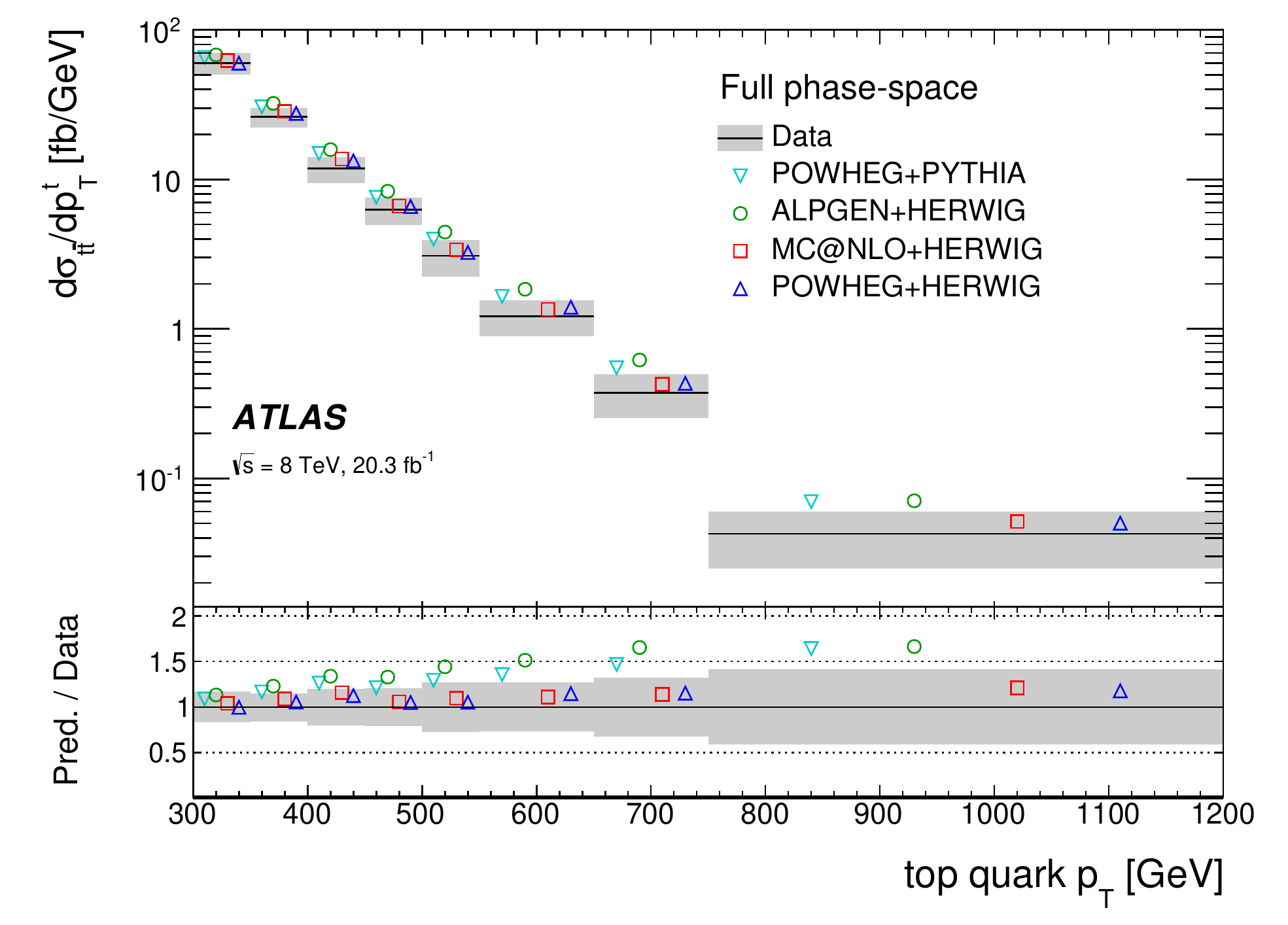} 
\caption{ (left) Top quark pair-production cross-section as a function of the centre-of-mass energy compared to the NNLO QCD calculation complemented with NNLL resummation (top++2.0). (middle) Fiducial particle-level differential cross-section as a function of the hadronic top-jet candidate transverse momentum. (right) Parton-level differential cross-section as a function of the hadronically decaying top quark transverse momentum. The measurements are compared to various Monte Carlo predictions normalized to the NNLO+NNLL inclusive cross-section of 253 pb.
Taken from Refs.~\cite{ATLAS-TopSummary,ATLAS-TopDiffXsec}.
\label{fig:topX}}
\end{figure}

Electroweak single top production goes through via the $Wtb$ vertex. The rare s-channel process is very challenging experimentally. All ATLAS measurements~\cite{ATLAS-TopSummary}, summarised in Figure~\ref{fig:rareTop}, show good agreement with theoretical predictions. The rare process of top quark pair-production in association with a vector boson allows to measure the $t\bar{t}V$ coupling directly. A fit to the data of the three studied final states with opposite-sign di-leptons, same-same di-leptons and tri-leptons considering $t\bar{t}W$ and $t\bar{t}Z$ processes simultaneously yields a significance of 5$\sigma$ (4.2$\sigma$) over the background-only hypothesis for $t\bar{t}W$ ($t\bar{t}Z$) production~\cite{ATLAS-ttV}. The measured cross-sections are consistent with the SM (see Figure~\ref{fig:rareTop}). The $t\bar{t}\gamma$ process was first observed in Run 1 $\sqrt{s} = 7$~TeV ATLAS data with 5.3$\sigma$ significance~\cite{ATLAS-ttg} and its measured fiducial production cross-section of ($63 \pm 8 \mathrm{(stat)} ^{+17}_{-13} \mathrm{(syst)} \pm 1 \mathrm{(lumi)} $)~fb agrees well with the NLO prediction of $48 \pm 10$~fb. 

\begin{figure}[!ht]
\centering
\includegraphics[width=8.4cm]{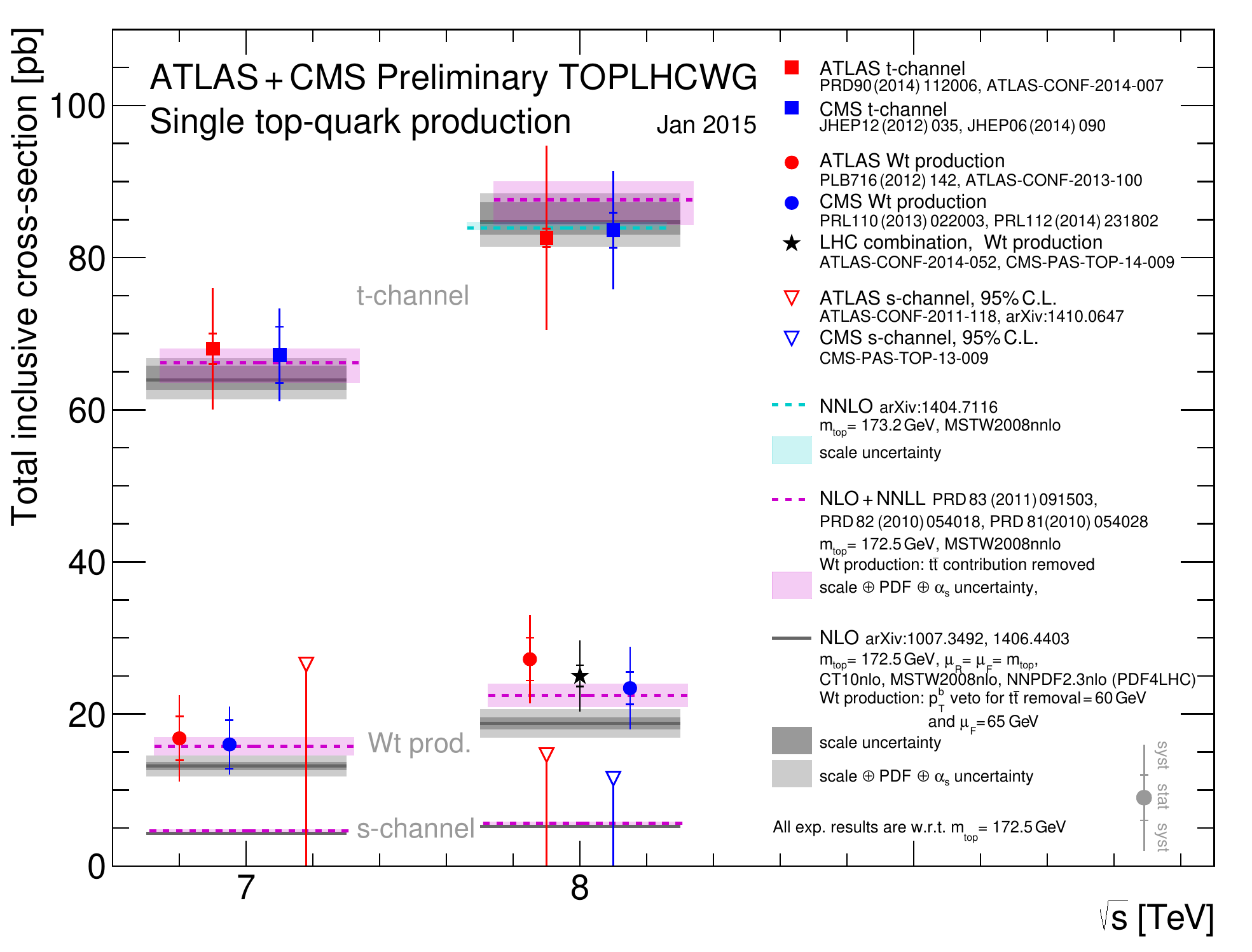} \hspace*{0cm}
\includegraphics[width=7.5cm, height=5.2cm,bb=0 35 570 390]{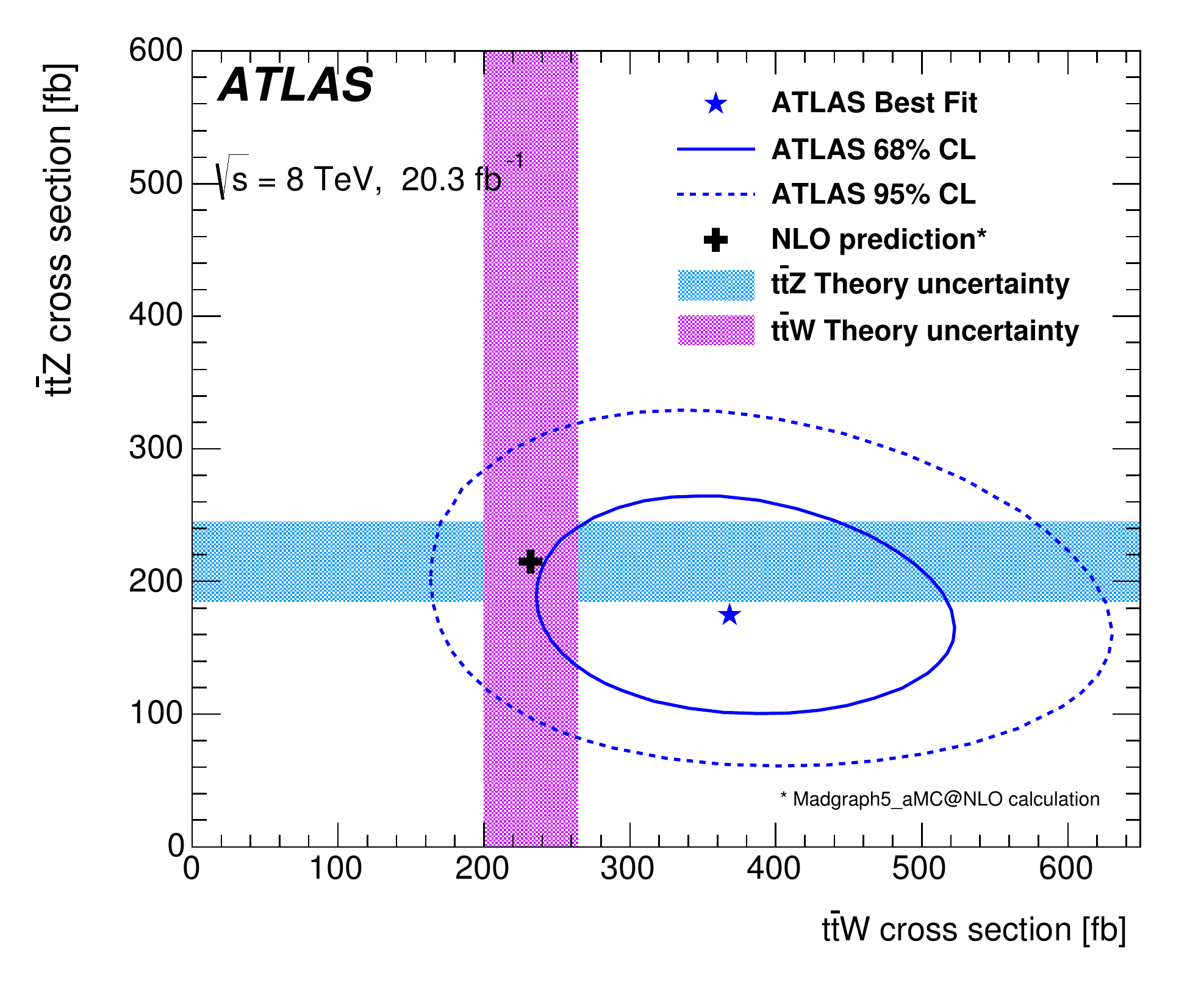}
\caption{ (left) Single top quark production cross-sections as a function of the centre-of-mass energy. (right) Simultaneous measurement of $t\bar{t}W$ and $t\bar{t}Z$ cross-section.
Taken from Refs.~\cite{ATLAS-TopSummary,ATLAS-ttV}.
\label{fig:rareTop}}
\end{figure}

\section{Beyond the Standard Model}

The open questions of the Standard Model motivated the birth of many theoretical models, including those featuring supersymmetry, extra dimensions, new forces mediated by heavy gauge bosons, strong dynamics leading to composite particles and more. ATLAS performs both general, (quasi) model-independent searches targeting different event topologies as well as exclusive ones aiming at the search of specific particles in a given process predicted by a chosen theoretical model.  
\subsection{Supersymmetry}

Supersymmetry (SUSY) is the most popular hypothesis to extend the SM. The high number of new parameters in its most general form calls for some reasonable assumptions to be made in order to arrive to a predictive theory that can experimentally be constrained. A comprehensive hunt is conducted at the LHC for the production of the supersymmetric partners of the elementary particles (light and 3rd generation squarks, gluinos, electroweakinos) with a huge number of models (cMSSM/mSUGRA, GMSB, pMSSM, simplified models...) and signatures covered, so far with no sign of new particles.

A summary of the many complementary searches for squarks and gluinos~\cite{ATLAS-MSUGRA} is shown in Figure~\ref{fig:SUSYIncl}. As visible from the plot, the measured Higgs mass also imposes constraints on the supersymmetric model parameters, in particular on the common scalar mass at the GUT scale ($m_0$). A representative selection of the various ATLAS search results illustrates the typical mass reach~\cite{ATLAS-SUSYSummary} on the right of Figure~\ref{fig:SUSYIncl}.

\begin{figure}[!ht]
\centering
\vspace*{-0.5cm}
\includegraphics[width=7.2cm,height=6cm, bb=20 10 600 400]{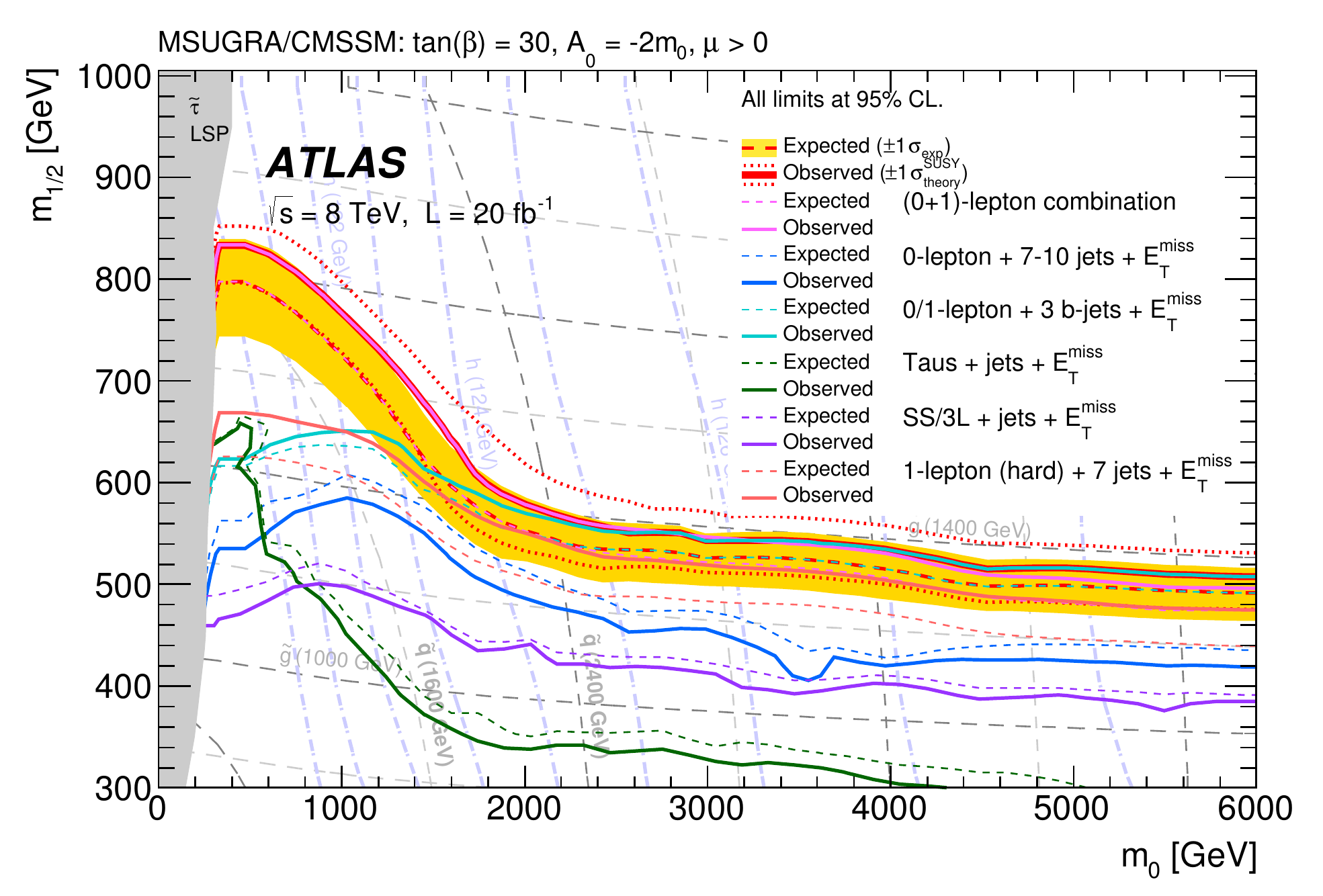} 
\includegraphics[width=9cm,height=8cm]{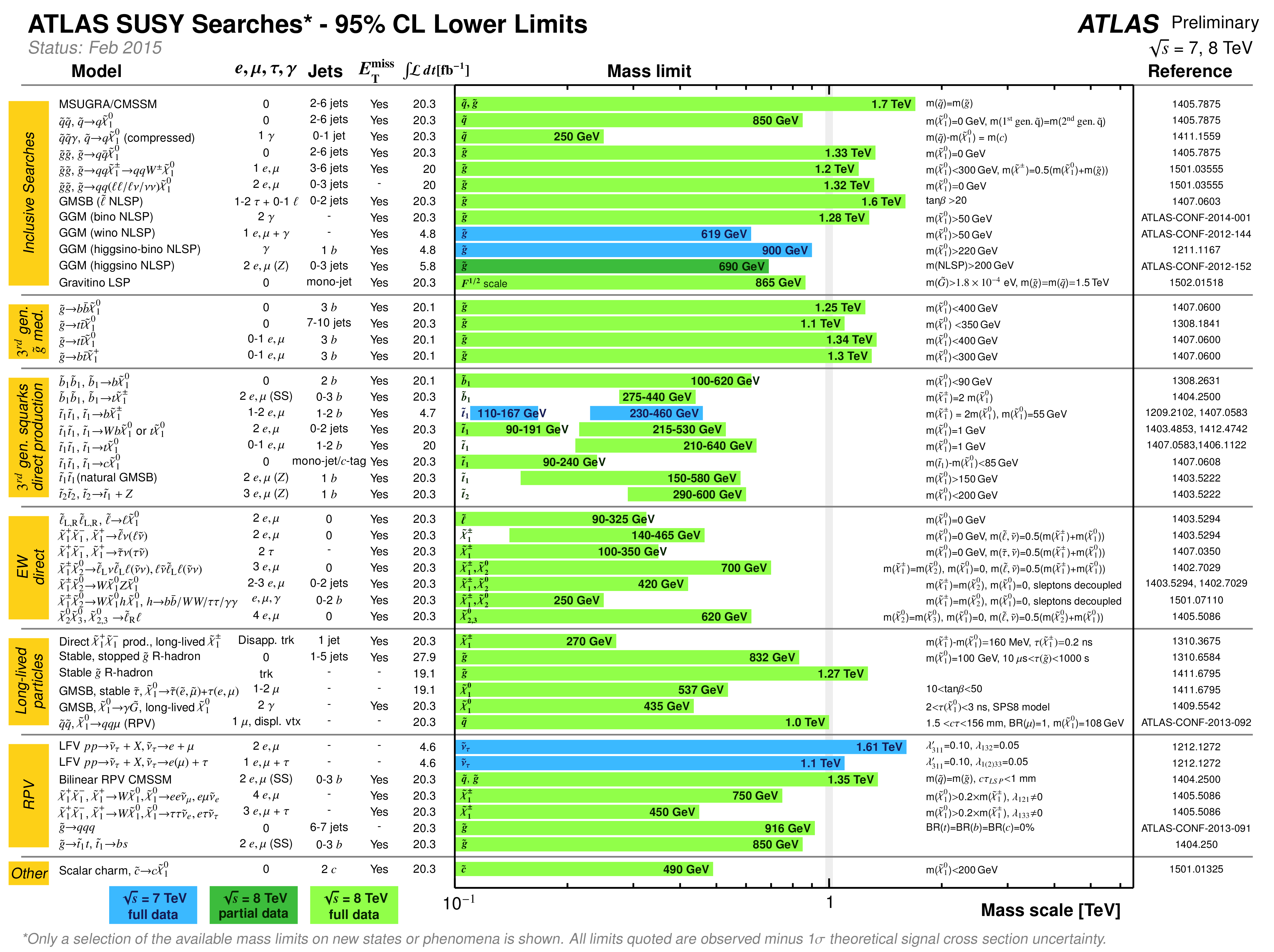} 
\caption{(left) Summary of inclusive searches for squarks and gluinos. Excluded regions are shown in the common scalar mass ($m_0$) -- common gaugino/higgsino mass ($m_{1/2}$) plane. (right)  Mass reach of a representative set of searches for supersymmetry.
Taken from Refs.~\cite{ATLAS-MSUGRA,ATLAS-SUSYSummary}.
\label{fig:SUSYIncl}}
\end{figure}

Measuring particle properties can also directly constrain New Physics. Such an example is the study of spin correlation of the top and anti-top quarks in $t\bar{t}$ events~\cite{ATLAS-TopSpin} (see Figure~\ref{fig:SUSYStop}). The measurement uses the azimuthal angle difference of the charged leptons in di-leptonic $t\bar{t}$ events to determine the degree of spin correlation with respect to the SM prediction. The measured distribution can also be used to perform a search for pair-production  of  top  squarks  with  masses close to the  top quark mass decaying to predominantly right-handed top quarks and a light neutralino, the lightest supersymmetric particle. Scalar top quark masses between the top quark mass and 191~GeV can be then excluded. This result complements traditional direct top squark searches and fills a part of the gap in the exclusion in the top squark mass - neutralino mass plane~\cite{ATLAS-Stop}  due to "stealth stop" regions where the SUSY signal has a topology very similar to the $t\bar{t}$ or $WW$ background. 

\begin{figure}[!hb]
\centering
\vspace*{-0.5cm}
\includegraphics[width=5cm,bb=20 0 560 600]{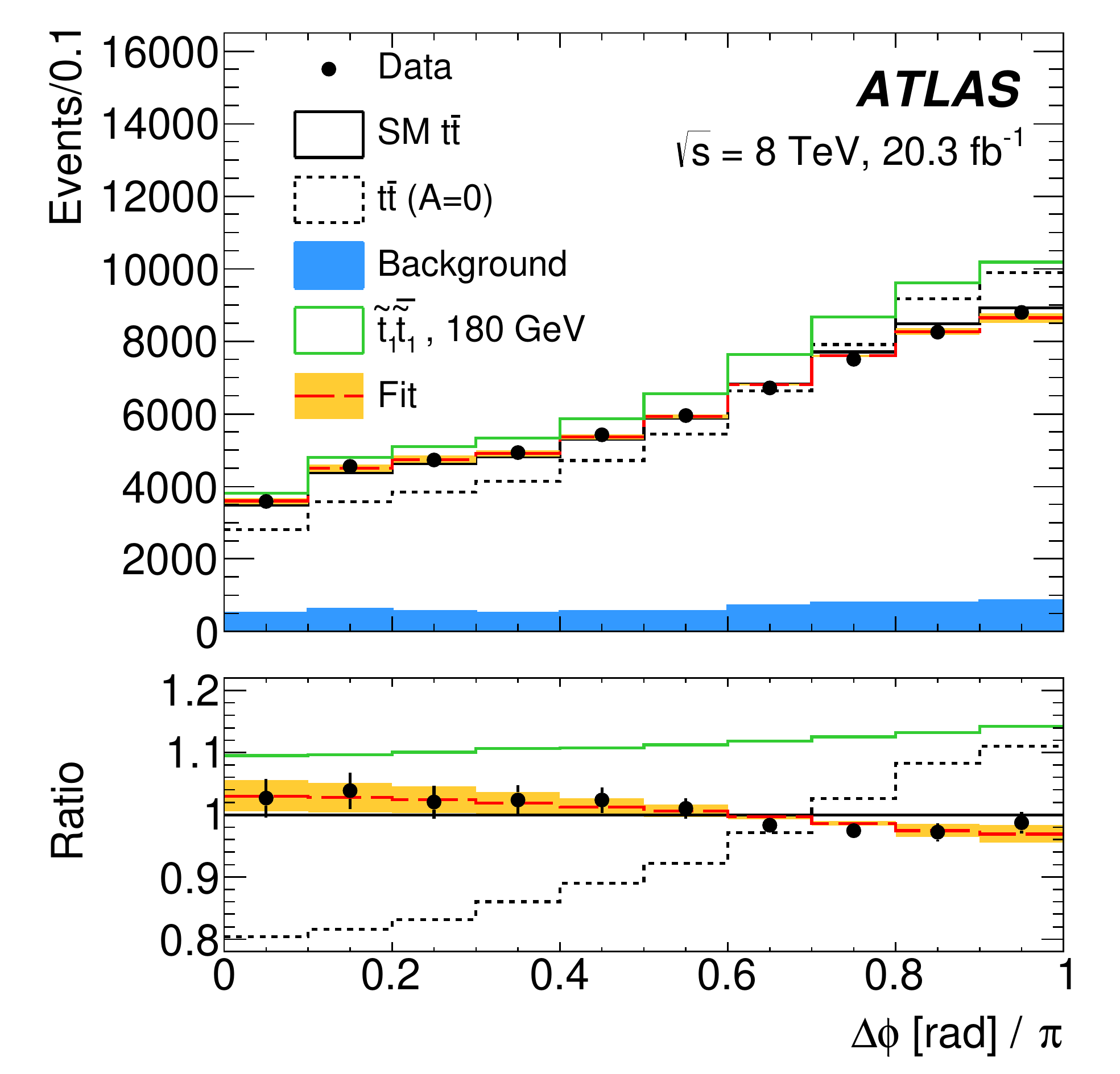} 
\includegraphics[width=5cm,bb=0 30 570 530,height=5cm]{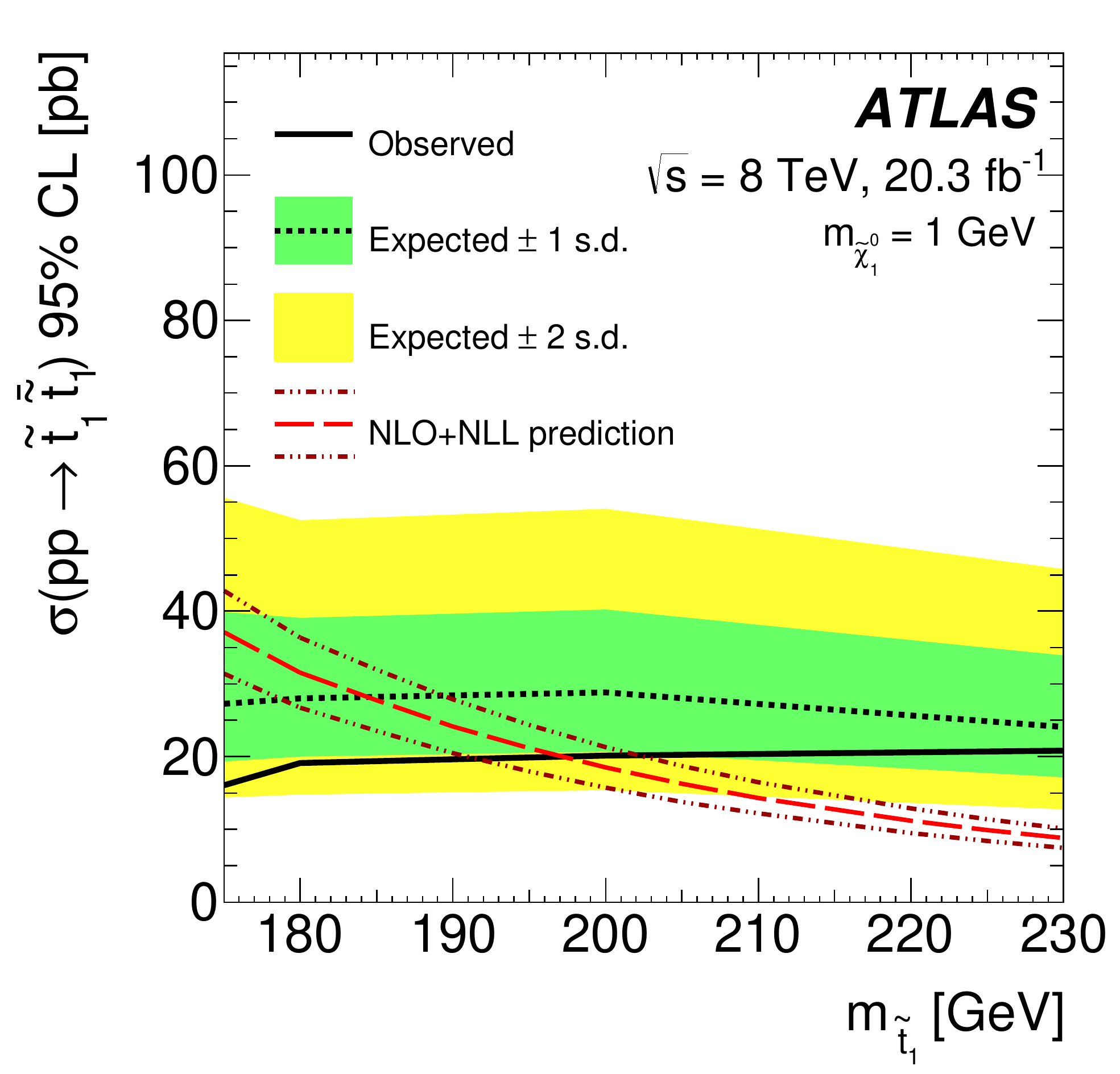} 
\includegraphics[width=6.2cm,bb=0 20 550 460,height=5.7cm]{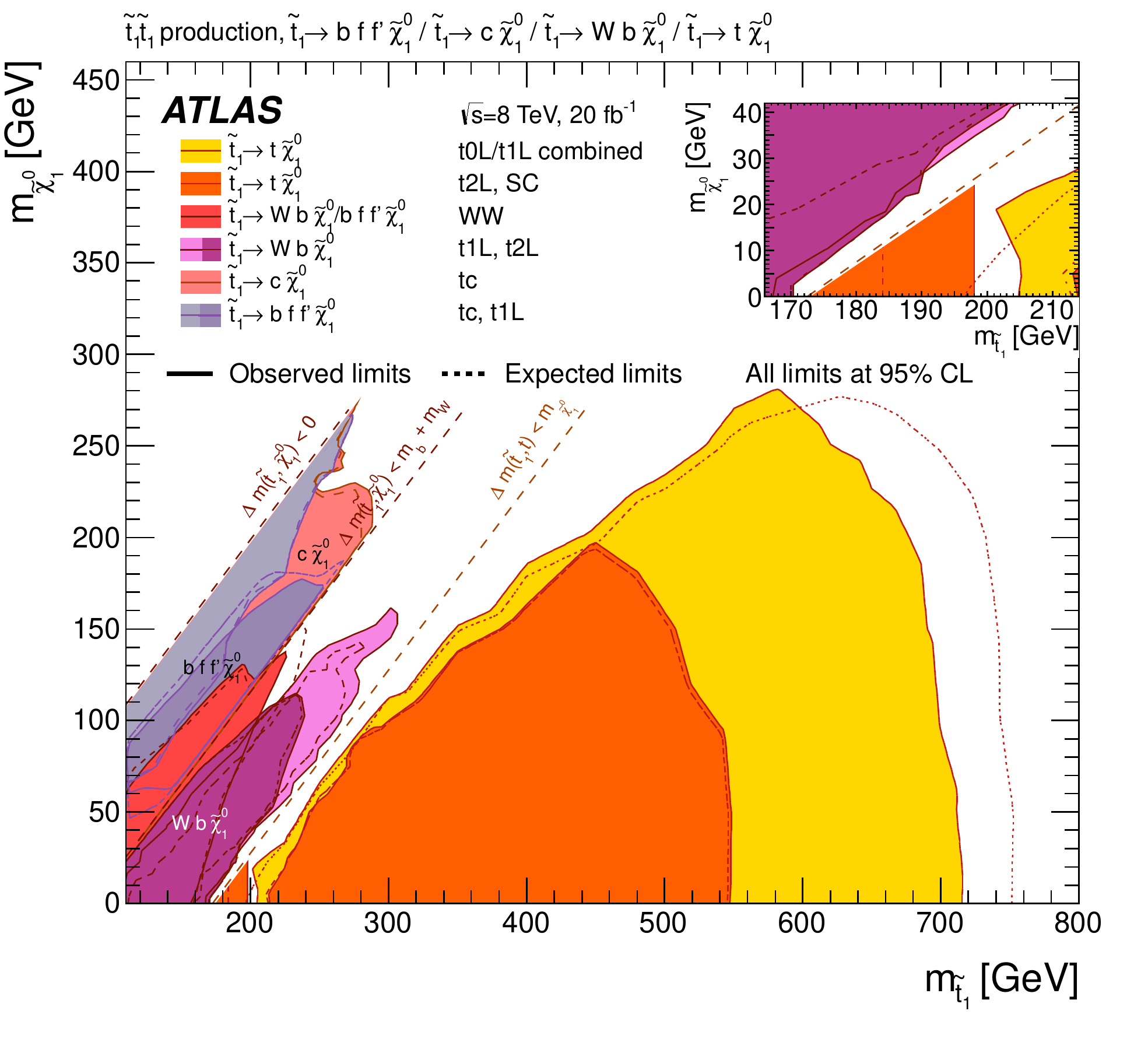} 
\caption{
(left) The distribution of the azimuthal angle difference between the charged leptons compared to the predictions of SM top quark pair-production and of the supersymmetric scalar top quark pair-production. The best fit to determine the degree of spin correlation with respect to the SM prediction is also shown. 
(middle) Expected and observed limits at 95\% CL on the top squark pair-production  cross-section  as  a  function of the stop mass for pair-produced top squarks decaying with 100\% branching ratio via $\tilde{t}_1 \rightarrow t\tilde{\chi}^0_1$ to predominantly right-handed top quarks, assuming a neutralino mas of 1~TeV.
(right) Summary of the dedicated ATLAS searches for top squark (stop) pair-production. The dashed and solid lines show the expected and observed limits, respectively, including all uncertainties except the theoretical signal cross-section uncertainty (PDF and scale). Four decay modes are considered separately with 100\% branching ratio. 
Taken from Refs.~\cite{ATLAS-TopSpin,ATLAS-Stop}.
\label{fig:SUSYStop}}
\end{figure}

In SUSY assuming the conservation of R-parity, a multiplicative quantum number that takes the value of +1 ($-1$) for SM (SUSY) particles, the lightest SUSY particle (LSP) is stable. For a neutral weakly interacting LSP, the characteristic signature will be of missing transverse energy (MET). A small but intriguing excess of events appeared in Run~1~\cite{ATLAS-SUSYZMET} in the final state with $Z \rightarrow \ell\ell$ accompanied by jets and MET (Z+jets+MET), as shown in Figure~\ref{fig:SUSYZMET}. Such a final state could be the result of gluino pair-production followed by the cascade decay $\tilde{g} \rightarrow \tilde{\chi}^0_1 q q \rightarrow \tilde{G} Z q q$, where the gravitino LSP would give rise to MET. The observed excess was 3.0$\sigma$ (1.7$\sigma$) in the $Z\rightarrow ee (\mu\mu)$ final state. 
This and other small observed excesses call for more data to disentangle whether they are statistical fluctuations or maybe the first signs of some New Physics.

\begin{figure}[!ht]
\centering
\includegraphics[width=6.5cm]{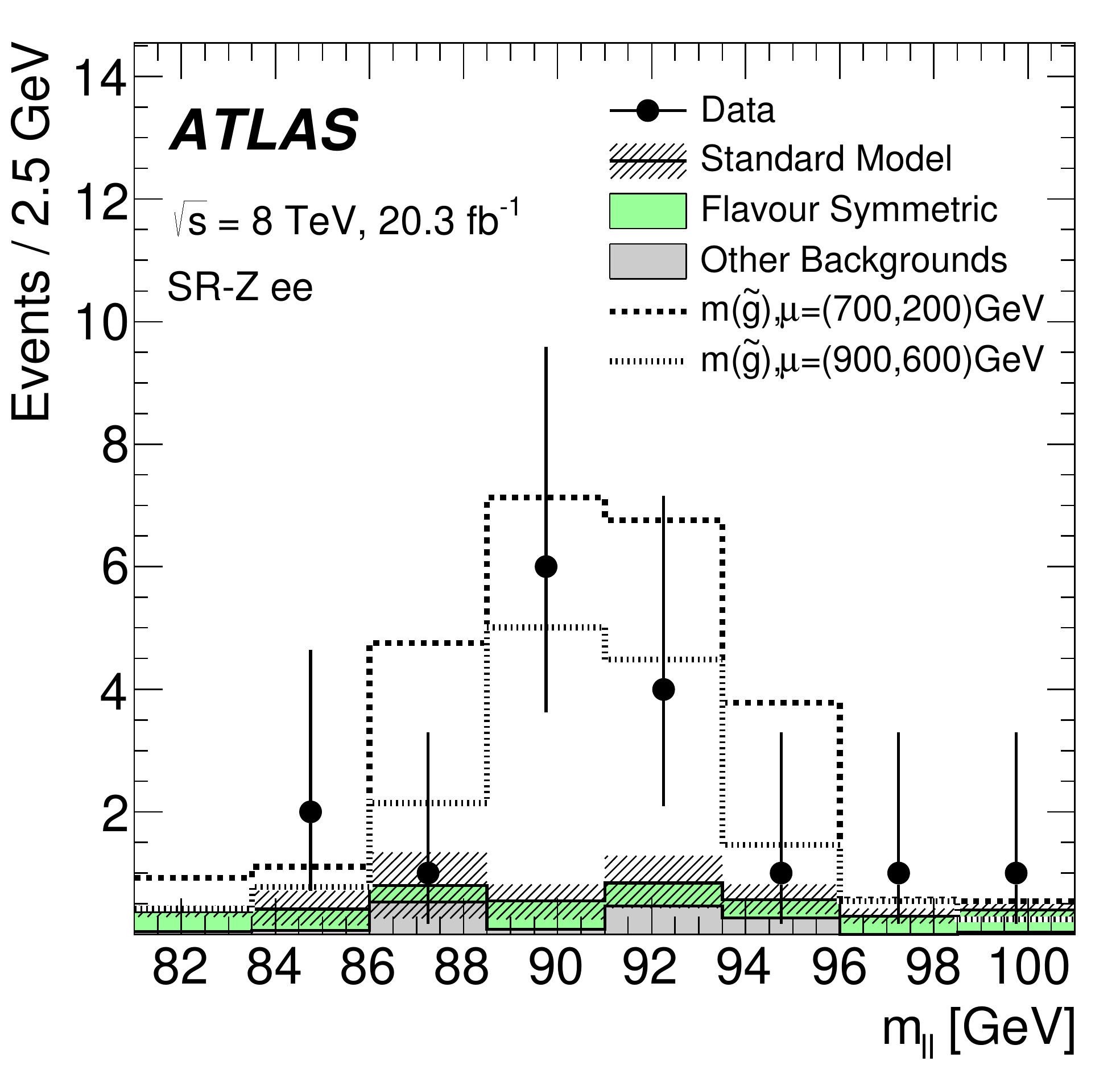} \hspace*{1cm}
\includegraphics[width=6.5cm]{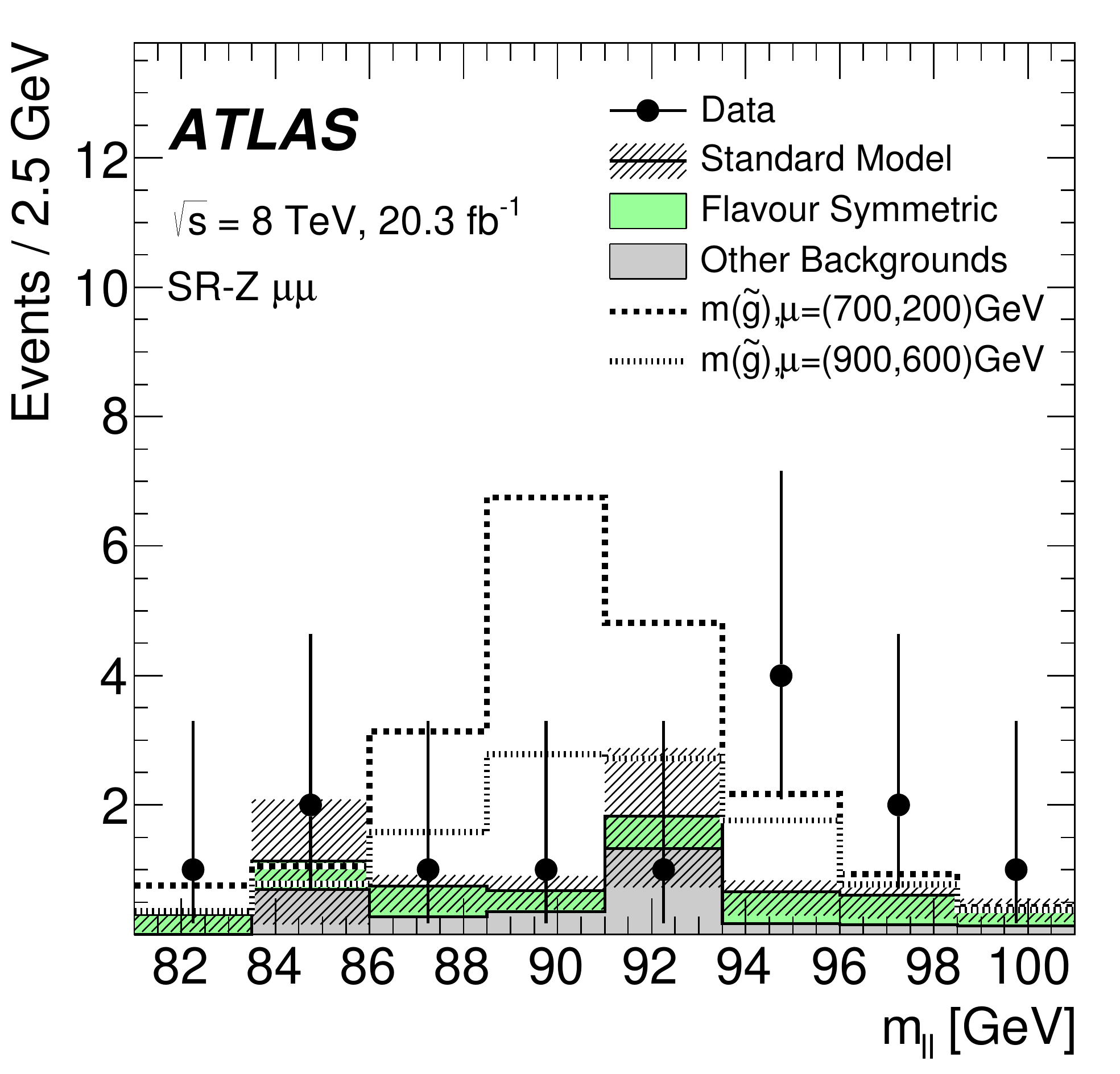} 
\caption{The di-lepton invariant mass distribution in the Z+jets+MET search in (left) the di-electron and (right) the di-muon channel.
Taken from Ref.~\cite{ATLAS-SUSYZMET}.
\label{fig:SUSYZMET}}
\end{figure}

\subsection{Exotic physics}

\begin{figure}[!hb]
\centering
\includegraphics[width=16cm, height=10.7cm]{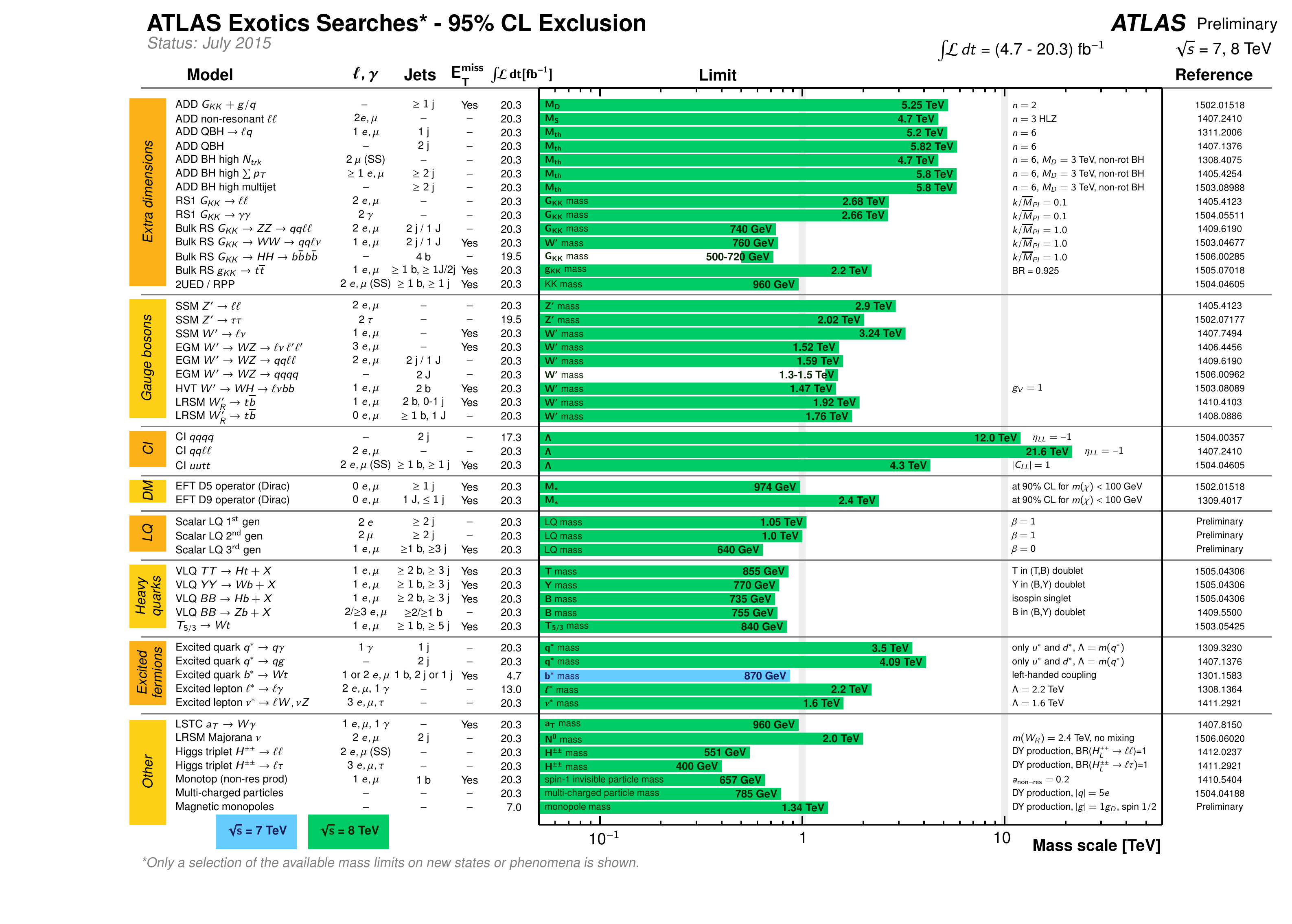}
\caption{Reach of ATLAS searches for new phenomena other than supersymmetry. Only a representative selection of the available results is shown.
Taken from Ref.~\cite{ATLAS-ExoSummary}.
\label{fig:ExoticsSummary}}
\end{figure}

New Physics can be of many types thus searches for non-supersymmetric phenomena are also a rich field under exploration. Many such searches~\cite{ATLAS-ExoSummary} are conducted and Figure~\ref{fig:ExoticsSummary} shows the results of a selection of them, including searches for Kaluza-Klein excitations of SM particles in theories with extra dimensions, for new gauge bosons, leptoquarks, vector-like heavy quarks, excited fermions and more. 

The most exciting result in Run~1 was the search for heavy bosons (such as a W' or a Kaluza-Klein graviton) decaying to a pair of weak vector bosons~\cite{ATLAS-WpJJ}. In the search where the resulting vector bosons are assumed to be boosted and to decay hadronically, each giving a fat jet with substructure, an excess of events was observed at a di-jet invariant mass of about 2~TeV as shown in Figure~\ref{fig:WZ}. This excess, when interpreted as coming from the process $W' \rightarrow WZ$, reaches a local (global) significance of $3.4\sigma$  ($2.5\sigma$). However other search channels does not confirm the result~\cite{ATLAS-OtherWp}.

\begin{figure}[!hb]
\centering
\includegraphics[width=7cm, height=6cm]{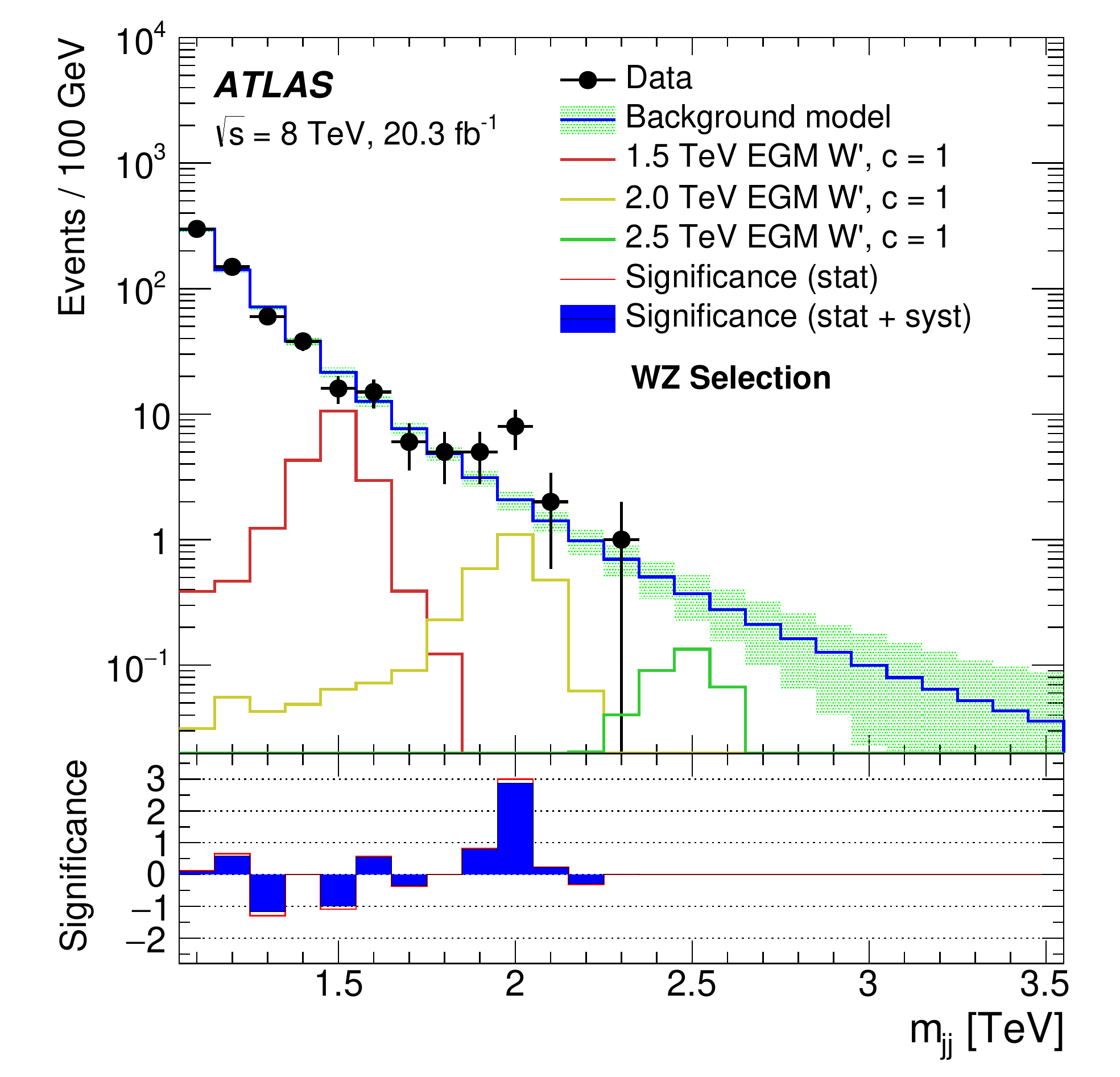} \hspace*{1cm}
\includegraphics[width=7cm, height=6cm]{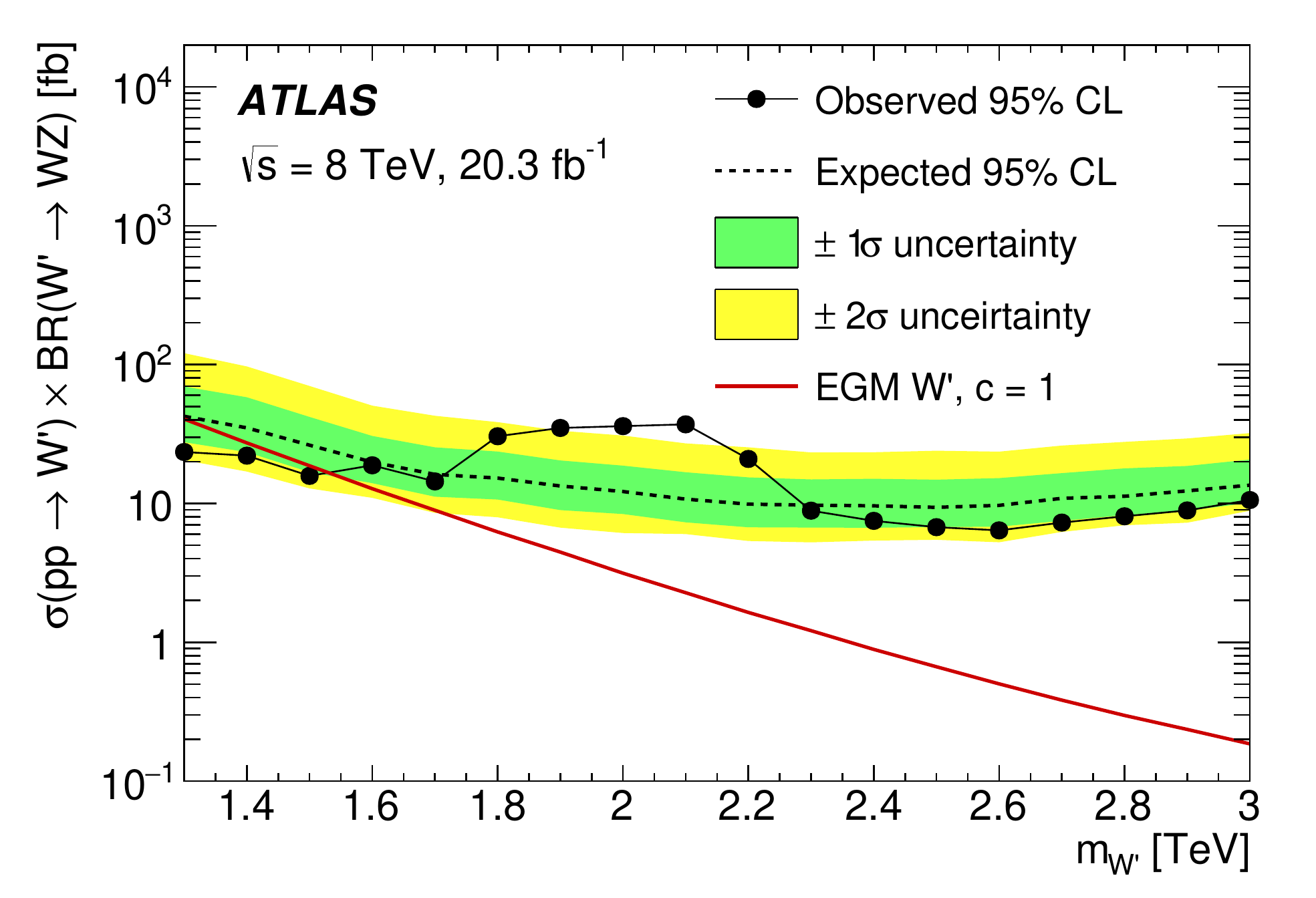} 
\caption{ (left) The reconstructed di-jet invariant mass in the search for $W' \rightarrow WZ \rightarrow jj$ process. (right) The excluded cross-section times branching ratio values as a function of the hypothetical W' mass.
Taken from Ref.~\cite{ATLAS-WpJJ}.
\label{fig:WZ}}
\end{figure}

\section{Summary and outlook}

The ATLAS detector performed very well both in LHC Run 1 and in 2015, in the first year of Run 2. In the latter period the upgraded detector was commissioned in a timely manner and new results appeared already in the summer of 2015. The about $\int\cal{L}$ = 25~fb$^{-1}$ Run 1 data provided a very rich harvest of new results, including the discovery and the subsequent measurements of the Higgs boson. While the Standard Model is now completed, the quest to discover the most fundamental rules that govern our Universe is as exciting as ever. In this spirit, Run 2 promises high-precision measurements and an improved sensitivity to New Physics. The first results using the 2015 data at $\sqrt{s} = 13$~TeV corresponding to almost $\int\cal{L}$ = 4~fb$^{-1}$ are in the limelight now~\cite{ATLAS-2015}. They allow the first measurements of for example W, Z and top quark production at the highest energies to date and extended in several searches the sensitivity for new phenomena.
The observed small excesses are begging for more data. The LHC is expected to deliver about $\int\cal{L}$ = 30~fb$^{-1}$ data in 2016, for the analysis of which ATLAS eagerly prepares.

\bibliographystyle{aipproc}

}